**Title**: Switching of Perpendicular Magnetization by Spin-Orbit Torque

**Author**: Lijun Zhu

State Key Laboratory of Superlattices and Microstructures
Institute of Semiconductors, Chinese Academy of Sciences, Beijing 100083, China
Email: ljzhu@semi.ac.cn

College of Materials Science and Opto-Electronic Technology
University of Chinese Academy of Sciences, Beijing 100049, China

**Abstract**: Magnetic materials with strong perpendicular magnetic anisotropy are of great interest for the development of nonvolatile magnetic memory and computing technologies due to their high stabilities at the nanoscale. However, electrical switching of such perpendicular magnetization in an energy-efficient, deterministic, scalable manner has remained a big challenge. This problem has recently attracted enormous efforts in the field of spintronics. Here, we review recent advances and challenges in the understanding of the electrical generation of spin currents, the switching mechanisms and the switching strategies of perpendicular magnetization, the switching current density by spin-orbit torque of transverse spins, the choice of perpendicular magnetic materials, and summarize the progress in prototype perpendicular SOT memory and logic devices toward the goal of energy-efficient, dense, fast perpendicular spin-orbit torque applications.
**Keywords:** Spin-orbit torque, magnetic anisotropy, spin polarization, spin current, magnetization switching

## 1. Introduction

"Memory wall" has become a critical bottleneck of information science in the post-Moore era.[1] As shown in **Figure** 1, while the feature size of the Intel CPU, a representative, keeps scaling down, the clock speed has almost stopped improving since 2014. A major reason is that the present memory technology cannot afford further performance enhancement of the CPU. It has become a consensus that many key electronic technologies, e.g., large-scale computing, machine learning, and superconducting electronics, would benefit from the development of new fast, non-volatile, and energy-efficient magnetic memory. Magnetic materials with strong perpendicular magnetic anisotropy (PMA) have the potential to be made into nonvolatile data bits with good scalability and high stability. However, electrical switching of such perpendicular magnetization in an energy-efficient, deterministic, scalable manner has remained a challenge.

Since the discovery that an in-plane charge current in certain heavy metal thin films can be utilized to manipulate and excite the magnetization state of an adjacent ferromagnetic layer,[2–9] spin-orbit torques (SOTs) have become a powerful and versatile tool to manipulate magnetic materials. SOTs are exerted when angular momentum is transferred from a spin current or spin accumulation. As shown in **Figure** 2a, a spin current with spin polarization $\sigma$ can exert two types of SOTs on a magnetization $M$, i.e., the damping-like torque [$\tau_{DL} \sim M \times (M \times \sigma)$] due to the absorption of the spin current component transverse to $M$ and field-like torque [$\tau_{FL} \sim (M \times \sigma)$] due to reflection of the spin current with some spin rotation. The same physics can be expressed in terms of effective SOT fields: a damping-like effective SOT field ($H_{DL}$) parallel to $M \times \sigma$ and a field-like effective SOT field ($H_{FL}$) parallel to $\sigma$.

Spin-orbit torques have attracted considerable attention in the field of magnetism and spintronics due to their great potential in nonvolatile memory and computing applications. Magnetic random access memory (MRAM) driven by spin-orbit torque, known as SOT-MRAM,[6,7,10-12] has the potential to mitigate the issues of conventional spin transfer torque (STT)-MRAMs (Figure 2b). In the latter, the required high write current limits the scaling and can exert severe stress on the magnetic tunnel junction (MTJ) and induce wear-out and breakdown of the tunnel barrier.[13] In a SOT-MRAM (Figure 2c), the spin current generated by the SHE of a heavy metal layer switches the magnetic free layer of a MTJ so that the read and write paths are separated, and writing requires only low voltages. The SOT-MRAMs can have long data retention, zero standby power, and fast writing.[10-14] Perpendicular SOT-MRAMs are also of particular interest because a strong PMA of the magnetic free layer can benefit the thermal stability in the tens of nanometer scale.

Since there have been very good reviews on perpendicular anisotropy,[15] spin-orbit torques,[16-21,22,23] domain wall motion,[24] and other more generalized aspects,[25] this review will focus on recent advances in the understanding of the electrical generation of spin current, the switching mechanisms and the switching strategies of perpendicular magnetization, the switching current density by the spin-orbit torques of transverse spins, choice of perpendicular magnetic materials, perpendicular SOT memory devices, perpendicular SOT logic devices, and finally conclusion and perspective.

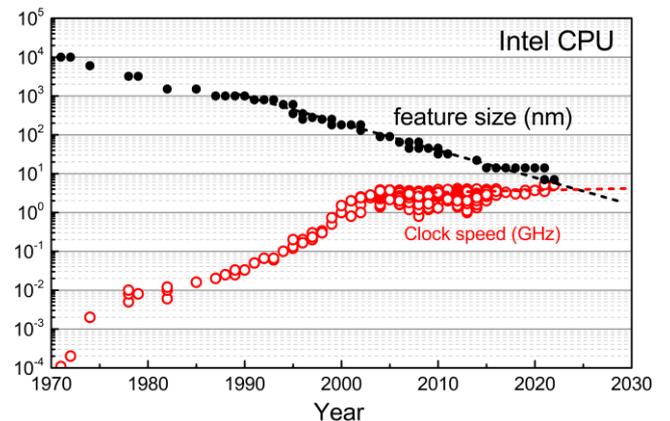

**Figure 1**. Clock speed and feature size of Intel CPU. The raw data is collected from https://ark.intel.com.

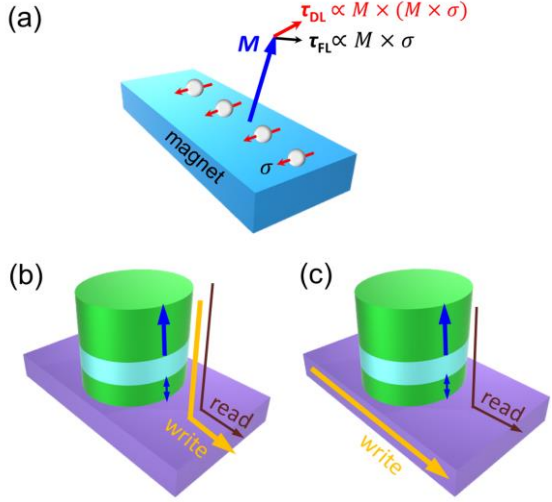

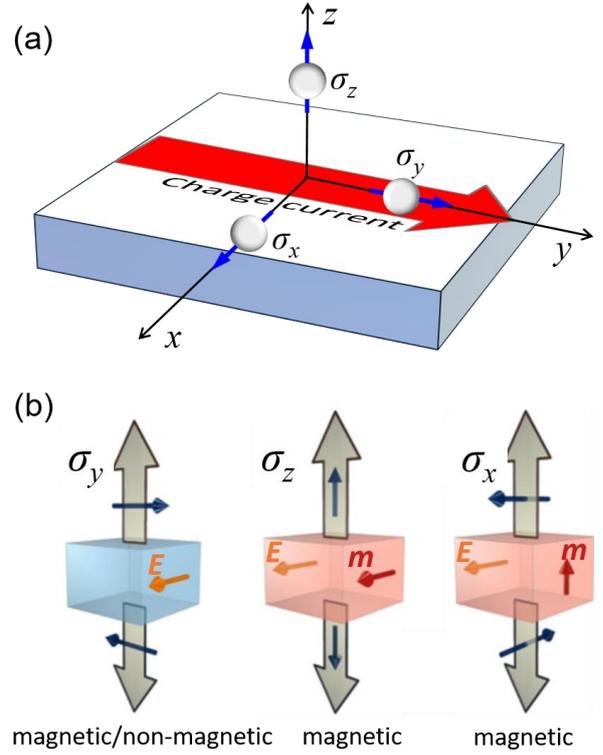

**Figure 2.** a) Damping-like and field-like spin-orbit torques exerted on a thin-film magnet *M* by a spin current with the polarization *σ*. b) Schematic of a perpendicular STT-MRAM, indicating that both the write and read currents flow through the magnetic tunnel junction. c) Schematic of a perpendicular SOT-MRAM, showing that the write and read currents flow separately through the spin Hall channel and the magnetic tunnel junction, respectively.

## 2. Electrical generation of spin currents

In general, the polarization of a spin current is a vector with three orthogonal components. As schematically shown in **Figure 3**a, a charge current flowing in the *x* direction could induce the generation of transverse spins ($\sigma_y$), perpendicular spins ($\sigma_z$), and longitudinal spins ($\sigma_x$) in presence of specific symmetry breaking in the material (see Davidson et al.[26] for an extended discussion). The Curie principle [27] allows for or prohibits possible system responses based on symmetries. It requires that the symmetry of an effect (e.g., electrical spin current generation) must coincide with the symmetry of the cause (e.g., the symmetries of the material and applied electric field). If the effect breaks a symmetry that the cause preserves, the effect cannot exist. If the cause breaks a symmetry, any effect that breaks the same symmetry is permissible.

Transversely polarized spin current or spin accumulation can be generated by a longitudinal charge current flow in either magnetic or non-magnetic materials (Figure 3b) via a variety of possible spin-orbit-coupling (SOC) effects. The latter includes the bulk spin Hall effect (SHE), [28-31] topological surface states,[32,33] interfacial SOC effects, [34-38] orbit-spin conversion, [39] the anomalous Hall effect, [40] the planar Hall effect,[41,42] the magnetic SHE,[41-45] Dresselhaus effect,[46] etc. The bulk SHE has been widely observed in thin-film heavy metal (HM), [4-12,47-49] Bi-Sb, [50] $Bi_xTe_{1-x}$,[51] CoPt,[52] FePt,[53] $Fe_xTb_{1-x}$,[54] and Co-Ni-B.[55] Generation of transverse spins by the SHE can be highly efficient as indicated by the giant interfacial damping-like SOT efficiency of $\geq 0.25$.[16,48,49-52,56] Highly efficient generation of transverse spins by the intrinsic SHE of heavy metals has enabled the demonstration of ultrafast, low-power, reliable anti-damping torque switching of MTJs by current pulses of 200 picosecond duration[11,57] and critical current density of $3.6\times10^6$ A/cm$^2$. [58]

**Figure 3.** Spin polarizations. a) Three orthogonal polarizations of spin currents: transverse spins ($\sigma_y$), perpendicular spins ($\sigma_z$), and longitudinal spins ($\sigma_x$). b) Symmetries that allow the perpendicular spin current flow with different polarizations in centrosymmetric materials. Transverse spins are allowed regardless of the magnetic configuration, while perpendicular (longitudinal) spins are allowed only in magnetic materials with magnetization along the longitudinal (perpendicular) direction.

Perpendicular and longitudinal spins are not allowed in nonmagnetic materials that are cubic crystal or polycrystalline/amorphous. For magnetic materials, perpendicular spins are allowed only when the non-zero magnetization is aligned in the longitudinal direction (Figure 3b). However, additional crystal or magnetic symmetry breaking can be introduced to make perpendicular and longitudinal spins permissive. So far, generation of perpendicular spins has been claimed in low-symmetry crystals (e.g., $WTe_2$,[59] $MoTe_2$,[60] and CuPt[61]), non-collinear antiferromagnetic crystals with magnetic asymmetry (e.g., $IrMn_3$,[62] $Mn_3GaN$,[63,64] $Mn_3Sn$ [65]), and also some collinear antiferromagnets with spin conversions (e.g., $Mn_2Au$).[66,67] Note that an in-plane charge current in a perpendicularly magnetized ferromagnet, which has different conductivities for the majority and minority carriers, can generate perpendicular spins flowing in the current direction (as in the case of STT devices) but not in the perpendicular direction required for SOT applications. Care is also needed to conclude the generation of perpendicular spins since non-uniform current effects[68] can also exhibit the characteristics that have been widely assumed to "signify" the presence of a flow of perpendicular spins, i.e., a sin2*φ*-dependent contribution in spin-torque ferromagnetic resonance (ST-FMR) signal of in-plane magnetization (*φ* is the angle of the external magnetic field with respect to the current), a φ-independent but field-dependent contribution in the second harmonic Hall voltage of in-plane magnetization, and external-field-free current switching of perpendicular magnetization.

Longitudinal spins are allowed by the symmetry when the non-zero magnetization is aligned in the perpendicular direction (Figure 3b), which has been experimentally discussed.[69,70] In addition, the generation of longitudinal spins has also been claimed from lateral spin valve experiments on low-symmetry crystals (*e.g.*, $MoTe_2$[71]) and from ST-FMR experiments on low-symmetry single-crystalline films and interfaces (e.g., (Ga,Mn)As,[72] NiMnSb,[73] and Fe/GaAs[74]).

## 3. Switching mechanisms of perpendicular magnetization

In the macrospin limit, a perpendicular magnetization behaves as a single, uniform magnetic domain and can be reversed coherently by an external magnetic field or a spin-orbit torque effective field that is greater than the PMA field ($H_k$). Coherent rotation of macrospin should follow the Stoner-Wohlfarth behavior [75] and have the following characteristics:

(i) the out-of-plane magnetic moment ($M_z$) and the anomalous Hall resistance have perfectly sharp reversal under a swept perpendicular magnetic field (see the red curves in Figure 4a,b);

(ii) the in-plane magnetic moment ($M_{xy}$) varies in linear proportion to an applied in-plane magnetic field ($H_{xy}$) that is not greater than $H_k$ and saturates for $|H_{xy}| > H_k$ (see the blue curve in Figure 4a);

(iii) the Hall resistance ($R_H$) should vary with the polar angle of the magnetization and thus $H_{xy}$ following

$$R_H = R_{AH} \cos(\arcsin(H_{xy}/H_k)). \tag{1}$$

As a result, $R_H$ is a parabolic function of $H_{xy}$ when $|H_{xy}| \ll H_k$ (see the blue curve in Figure 4b).

(iv) the coercivity ($H_c$) of the perpendicular magnetization should vary with the polar angle of the driving magnetic field ($\theta_H$) following

$$H_c = H_k (\cos^{2/3}\theta_H + \sin^{2/3}\theta_H)^{-3/2}, \tag{2}$$

When the magnetic field is applied along the film normal, i.e., $\theta_H = 0°$ or $180°$, the switching field $H_c$ of a perpendicular macrospin is equal to its anisotropy field $H_k$.

Switching of a multi-domain magnet is mainly via reversed domain nucleation and domain wall propagation because this requires much less energy than switching via coherent rotation. In this case, the aforementioned macrospin behaviors are disobeyed in part or in full. For example, $H_c$ would represent the depinning field of the magnetic domain walls rather than the magnetic anisotropy field. $H_c$ follows a $1/\cos\theta_H$ scaling [76,77] and is typically much smaller than the perpendicular magnetic anisotropy field at $\theta_H = 0°$ or $180°$.

A ferro- or ferrimagnetic material becomes a macrospin after it is magnetized by a sufficiently large magnetic field along its magnetic easy axis, which allows for the application of macrospin approximation in low-field harmonic Hall voltage measurements, [78-81] etc. However, near and during switching most perpendicular magnetic anisotropy samples transition into multi-domain configuration and show reversed domain nucleation and domain wall propagation (see Kumar *et al.*,[24] for an extended discussion). This can be seen from polar MOKE images during switching, see examples of Pt 5 nm/Co 0.6 nm [82] in **Figure 5**a and of Pt 3nm/Co 1nm/Ru 1 nm/FeTb 6 nm [83] in Figure 5b.

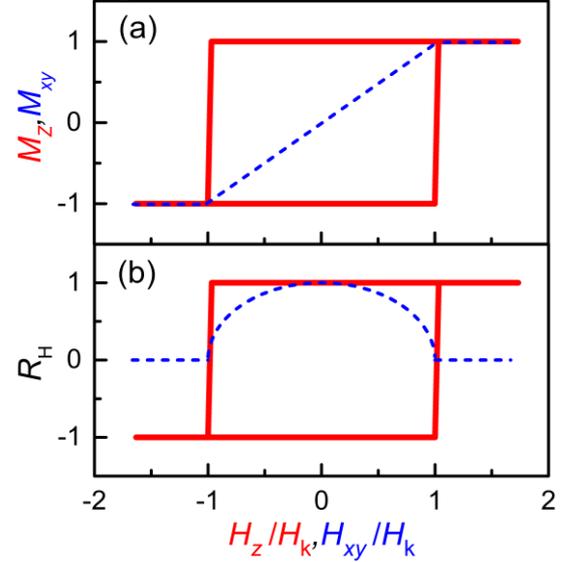

**Figure 4**. Schematic of Stoner-Wohlfarth macrospin behaviors. a) Out-of-plane component of magnetic moment ($M_z$) vs the ratio of the applied perpendicular magnetic field ($H_z$) to the effective perpendicular magnetic anisotropy field ($H_k$), (red) and the in-plane component of magnetic moment ($M_{xy}$) vs the ratio of the in-plane magnetic field ($H_{xy}$) to $H_k$ (blue). b) Hall resistance ($R_H$) vs $H_z/H_k$ (red) and the Hall resistance ($R_H$) vs $H_{xy}/H_k$ (blue).

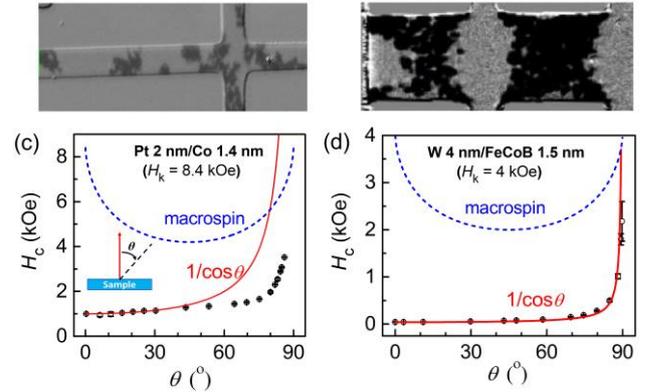

**Figure 5**. Polar MOKE images during magnetization switching of a) Pt 5 nm/Co 0.6 nm and b) Pt 3 nm/Co 1 nm/Ru 1 nm/FeTb 6 nm, showing reversed domain nucleation and domain wall propagation. Coercivity ($H_c$) vs the polar angle of the applied magnetic field ($\theta_H$) for c) Pt 2 nm/Co 1.4 nm and d) W 4 nm/FeCoB 1.5 nm, showing significant deviation from the expectation of macrospin [Equation (2), dashed blue line] and relatively good consistency with domain wall depinning model ($\propto 1/\cos\theta_H$, solid red line). The $H_c$ values are determined from the first harmonic Hall voltage hysteresis loops with the external field swept in the *xz* plane at different fixed polar angles of $\theta_H$. a) Reprinted under the terms of Creative Commons Attribution 4.0 International License.[82] Copyright 2018, Springer Nature. b) Reprinted with permission.[83] Copyright 2020, American Physical Society. c,d) Reprinted with permission.[84] Copyright 2021, American Physical Society.

As shown in Figure 5c,d, the magnetic field switching of the Pt/Co and W/FeCoB samples behavior distinctly from a macrospin and agrees better with the expectation of domain wall depinning. [84] The observed $H_c$ for both samples is much smaller than $H_k$ and significantly deviates from the scaling of Equation (2). Instead, $H_c$ much more closely follows a $1/\cos\theta_H$ scaling, which is predicted by the switching via thermally-assisted reversed domain nucleation and domain wall propagation. [76,77] For the Pt/Co sample, there is some deviation from the $1/\cos\theta_H$ behavior as $\theta_H$ approaches 90° likely due to coherent rotation of the magnetization vector in the pinned domain when the in-plane hard-axis field component is sufficiently strong. Such deviation is absent in the W 4 nm/FeCoB 1.5 nm sample with a weak pinning field ($H_c$ = 40 Oe), suggesting minimal magnetization rotation in this sample.[84] These observations are consistent with perpendicular MTJ experiments that the rigid macrospin reversal merely happens unless the device size is rather small (< 50 nm).[10,85]

## 4. Switching strategies of perpendicular magnetization

### 4.1 Switching by effective perpendicular magnetic field

So far, several strategies have been developed to switch perpendicular magnetization. Most obviously, perpendicular magnetization can be reversed by a perpendicular magnetic field that is greater than the coercivity of the magnet (the depinning field or perpendicular anisotropy field). Such a perpendicular magnetic field can be an externally applied magnetic field, an Oersted field of laterally asymmetric charge current channel, or a field-like SOT effective field of perpendicular spins. However, this method is of poor scalability and/or relatively low energy efficiency in switching large-coercivity magnetization.

### 4.2 Switching by perpendicular spins

In principle, a perpendicularly polarized spin current can manipulate perpendicular magnetization via antidamping spin torque in the absence of any external magnetic field. This would be the most ideal approach once high-density perpendicular spins be generated in an energy-efficient, high-endurance, and scalable manner.

Following the symmetry analysis (Figure 3b), perpendicular spins flowing in the z direction are allowed to be generated from the bulk or interface of an in-plane magnetization that is collinear with an in-plane current. This has been suggested to induce switching of the perpendicular magnetic component within the trilayers of in-plane CoFeB/Ti/perpendicular CoFeB (Figure 6a).[86,87] Meanwhile, when additional crystal or magnetic symmetry breaking is introduced, perpendicular and longitudinal spins can also be generated, e.g., in low-symmetry crystals [59-61], non-collinear antiferromagnetic crystals with magnetic asymmetry [62-65] or magnetic SHE,[65,88] and also some collinear antiferromagnet crystals with spin conversions [66,67] or spin splitting effect [89-91] (Figure 6b). However, this approach requires single-crystal structures, which is not friendly to the integration of such perpendicular-spin generators into semiconductor transistor circuits.

While it has remained an unmet challenge, highly efficient generation of perpendicular spins in polycrystalline systems, if somehow achieved in the future, would be a significant breakthrough towards the development of dense and low-power perpendicular SOT-MRAMs that can be switched reliably without any external field.

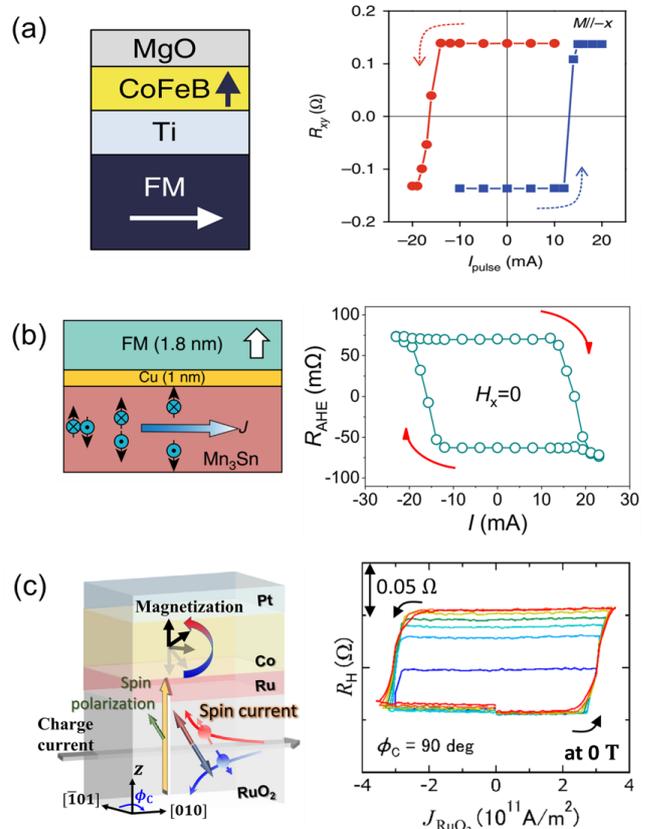

**Figure 6**. External-magnetic-field-free switching of perpendicular magnetization by an in-plane current in a) in-plane CoFeB (1nm)/Ti (3nm)/perpendicular CoFeB/MgO, b) Mn$_3$Sn (7nm)/Cu (1nm)/perpendicular CoFeB (1.8 nm), and c) RuO$_2$ (10 nm)/Ru (0.8 nm)/Co (0.8 nm)/Pt (2.0 nm). a) Reprinted with permission.[86] Copyright 2018, Springer Nature. b) Reprinted under a Creative Commons Attribution 4.0 International License.[88] Copyright 2022, Springer Nature. c) Reprinted with permission.[89] Copyright 2022, American Physical Society.

### 4.3 Switching by transverse spins

Transverse spins ($\sigma_y$) can manipulate perpendicular magnetization by overcoming the coercivity or perpendicular magnetization via the effective field ($\sim M \times \sigma_y$) of an interfacial [47-51] or bulk [52-54] damping-like SOT. However, switching of perpendicular magnetization by transverse spins generated by heavy metals requires a markedly high write current in the ns and sub-ns pulse regime [14,92] compared to the collinear in-plane SOT devices [12,58] because the orientation of the spin polarization of the spin current from heavy metals is in the plane rather than parallel to the perpendicular magnetic easy axis. This is particularly so considering that a highly resistive Ta or W channel that leads to substantial current shunting into the magnetic free layer is typically required to achieve perpendicular magnetic anisotropy of the magnetic free layer.[16] To benefit from other spin-current generating materials, such as the optimized low-resistivity Pt-alloys and multilayers [11,16,48,58,81,93-96] or topological insulators, [32,33] an additional insertion layer of proper thin nonmagnetic layer (made of Mo, Ti, Cr, etc.) between the spin Hall channel and the magnetic free layer is typically needed to promote the PMA for the SOT-MRAMs.

Furthermore, switching of perpendicular magnetization by transverse spins requires the assistance of a strong in-plane magnetic field along the write current direction, i.e., longitudinal assisting magnetic field. In the case of macrospin, the longitudinal assisting magnetic field breaks the reversal symmetry of a macrospin [5] such that selective switching is possible. For multidomain samples, which is often the case for studies in the literature, the longitudinal assisting field overcomes the Dzyaloshinskii–Moriya interaction (DMI) at the interface of the spin Hall channel and the magnetic free layer. [77] Without a longitudinal assisting field, the magnetic moments inside the domain walls typically orientate following the DMI field such that the SOT fields exerted by the spin current are opposite for the domain wall moments on the two sides, leading to the displacement of the domain without inflation (**Figure 7**a). When a sufficiently large in-plane magnetic field is applied to align the domain wall moments along the current direction, the anti-damping SOT field on the domain wall would be in the same direction that promotes the growth or shrink of the domain (Figure 7b). In contrast, no spin-orbit torque should be expected when the domain wall moments are transverse to the current and thus collinear with transverse spins due to a large transverse magnetic field ($H_y > H_{DMI}$).

So far, the required in-plane longitudinal assisting magnetic field is achievable by exchange bias field from an adjacent antiferromagnetic layer (Figure 7d), [97,98] stray field,[14,99] interlayer coupling, [100,111] or the Néel orange-peel effect [111,112] from a nearby ferromagnetic layer (Figure 7e,f), a built-in magnetic field from a lateral structural asymmetry (non-uniform oxidization of a wedge-shaped magnetic free layer (Figure 7g),[113] a wedge-shaped ferromagnetic layer (Figure 7h),[114] a wedge-shaped heavy metal (Figure 7i),[115-107] a wedge-shaped insertion layer (Figure 7j), 108 a composition-wedged magnetic free layer (Figure 7k),[109] asymmetric contacts,[68] etc.), a field-like spin-orbit torque of longitudinal spins,[69,70] a strained PMN-PT substrate, [110] or a spin transfer torque from additional large write current in the magnetic tunnel junction (MTJ) nanopillar.[111,112] Very recently, field-free partial magnetization switching by an in-plane current is reported in a few samples,[113-119] but not in others, [120-122] with an out-of-plane composition gradient for the mechanism yet to carefully specify.

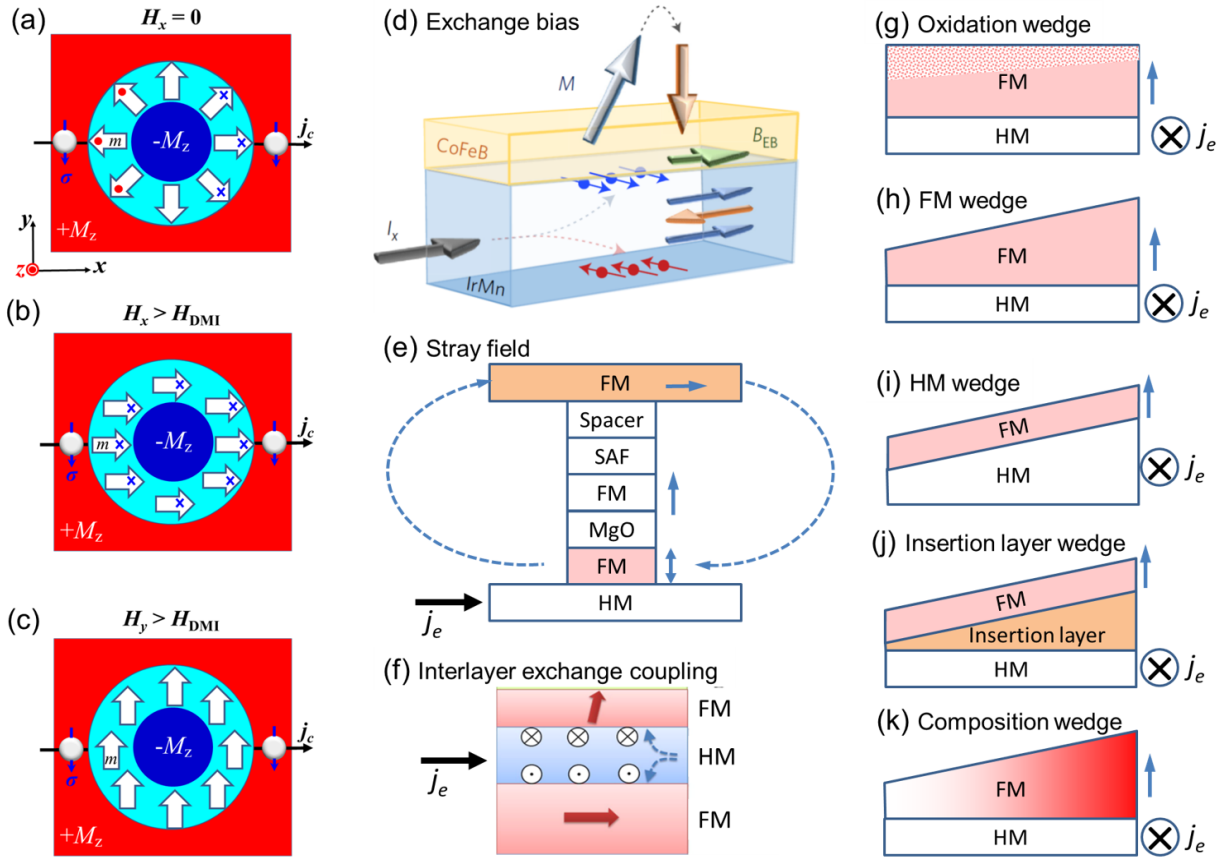

**Figure 7**. Role of magnetic fields during the spin-orbit torque switching process of perpendicular magnetization via domain wall motion. Damping-like SOT field on the magnetic moment inside of the magnetic domain wall in the case of a) no in-plane magnetic field, b) a longitudinal magnetic field greater than the DMI field ($H_x > H_{DMI}$), and c) a transverse magnetic field greater than the DMI field ($H_y > H_{DMI}$). Generation of effective in-plane assisting magnetic field in perpendicular magnetization by d) exchange bias field, e) stray field, f) interlayer exchange coupling, g) non-uniform oxidation, h) ferromagnet (FM) wedge, i) heavy-metal (HM) wedge, j) insertion wedge, and k) composition wedge. d) Reprinted with permission.[98] Copyright 2016, Springer Nature.

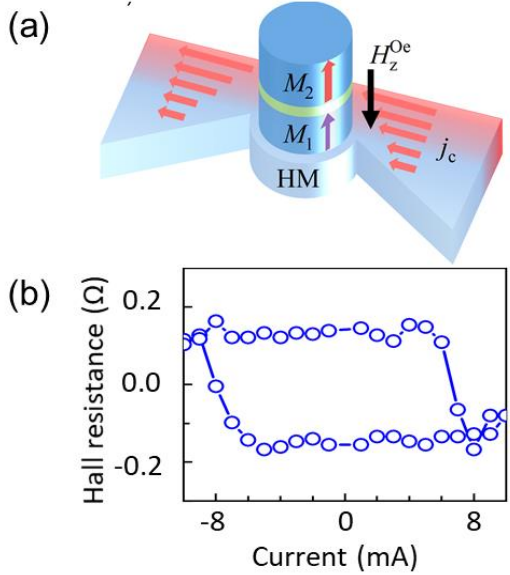

**Figure 8**. a) Schematic of a spin-orbit-torque magnetic tunnel junction with asymmetric current channel geometry. b) Switching of the Hall resistance for Pt (5 nm)/FeTb (7.5 nm) bilayer by an in-plane current in absence of an external magnetic field. Reprinted with permission.[68] Copyright 2022, AIP Publishing.

However, the effectiveness of all the above strategies requires experimental verification at the device level. The existing studies suggest that exchange bias field from a relatively thick adjacent antiferromagnetic layer (e.g., 8 nm IrMn) could lead to gradual switching,[97,98] which is detrimental to digital memory application that requires the MTJs to have very sharp, reliable switching between parallel to anti-parallel states. The lateral structural asymmetry strategies in Figure 7g,k can be hardly scalable in mass production. A strained PMN-PT substrate will challenge the integration with CMOS circuits due to its requirements of single-crystal structure and very high gate voltage for strain generation (500 V was applied to a perpendicular Hall bar device that requires only a small in-plane assistant field of 3 Oe in Ref. [110]). Inserting an in-plane magnetized metallic layer beneath the spin Hall channel to generate stray field or exchange interaction obviously leads to a significant waste of write current and thus lowers the energy efficiency of the SOT-MRAM devices (most magnetic insulators have weak magnetic anisotropy and small magnetization, which are not preferred for generating a strong stray field). The stray field generated by putting an in-plane hard magnetic layer on the top of the MTJs (Figure 7e) avoids wasting additional write current and thus is more energy-efficient, but the large dimension and stray field of the in-plane hard magnetic layer challenge the scalability.

Very recently, external-magnetic-field-free SOT switching of uniform perpendicular magnetization has also been achieved by the damping-like spin-orbit torque of transverse spins and an effective perpendicular magnetic field arising from current spreading near electrical contacts or asymmetric spin Hall channel of magnetic heterostructures [68] (**Figure 8**a). Utilizing such an effective perpendicular magnetic field, Liu and Zhu have demonstrated the electrical switching of polycrystalline Pt/FeTb bilayers with a giant perpendicular magnetic anisotropic field of 30 kOe and a large coercivity of 0.5 kOe in absence of an external magnetic field (Figure 8b). This strategy is technologically promising since it allows to switch a homogeneous perpendicular magnetization, e.g., in magnetic memory and logic, in a scalable, integration-friendly, and energy-efficient manner.

## 5. Current density for switching by transverse spins

### 5.1 Quantitative models of switching current density

For SOT-MRAMs (Figure 2c), the density of the critical switching current inside the spin-current generating layer ($j_{c0}$) is considered an important parameter. This is mainly because it is directly related to the *total* switching current ($I_c$) that will define the energy efficiency ($\propto I_c^2$), the scalability (the transistor dimension $\propto I_c$), and the endurance (electro-immigration $\propto I_c^2$) of SOT-MRAMs. It is important to note that the number and dimension of the transistors that source currents to generate the SOTs and to read the resistance are much more critical than the size of the MTJ nano-pillars in the determination of the scalability of SOT-MRAMs. As seen from the cross-sectional scanning electron microscopy (SEM) picture of a typical SOT-MRAM cell in **Figure 9**a, the transistor circuits and the electrodes are remarkably larger than the MTJ pillars. Quantitatively, the *total* switching current (Figure 9b) is the sum of the currents in the spin-current generating layer $I_{c0}$ and the ferromagnetic layer $I_{FL}=sI_{c0}$, i.e.,

$$I_c = (1+s)I_{c0} \qquad (3)$$

with $I_{c0} = j_{c0}wd$ and $s=\rho_{HM}t/\rho_{FL}d$. Here $w$, $d$, and $\rho_{HM}$ are the width, thickness, and resistivity of the spin Hall channel, $t$ and $\rho_{FL}$ are the thickness and resistivity of the magnetic free layer. Consequently, $j_{c0}$ is proportional to $I_c$ for a given device but not for varying devices.

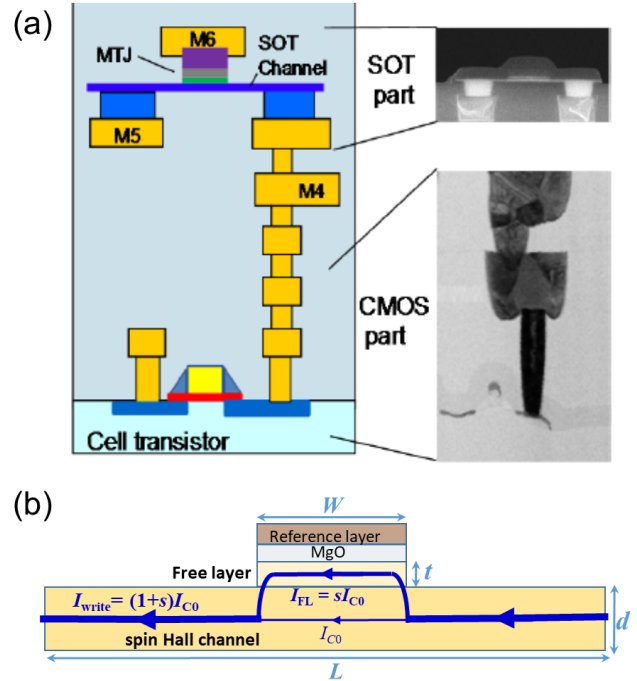

**Figure 9**. a) Cross-sectional schematic and cross-sectional SEM view of canted SOT MRAM cell. Reprinted with permission.[177] Copyright 2019, IEEE. b) Current shunting into the magnetic free layer.

In macrospin limit,[123,124] the transverse-spin-induced damping-like SOT efficiency per unit current density ($\xi_{DL}^{j}$) of a heterostructure with PMA inversely correlates to the critical switching current density ($j_{c0}$) in the spin-current-generating layer via

$$j_c = (e\mu_0 M_s t_{FM} H_k/\hbar \xi_{DL}^j)$$
$$\times \sqrt{1 + \frac{1}{8}[20(H_x/H_k)^2 - (H_x/H_k)^4 - (H_x/H_k)(8 + (H_x/H_k)^2)^{3/2}]}$$
(4)

or

$$j_c \approx e\mu_0 M_s t_{FM} (H_k - \sqrt{2}|H_x|)/\hbar \xi_{DL}^j, \quad (5)$$

where $e$ is the elementary charge, $\hbar$ is the reduced Planck constant, $\mu_0$ is the permeability of vacuum, $H_x$ is the applied field along the current direction, and $t_{FM}$, $M_s$, $H_k$, and $H_c$ are the thickness, the saturation magnetization, the effective perpendicular anisotropy field, and the perpendicular coercivity of the driven magnetic layer, respectively.

In the simplified domain wall depinning model, [125-127] $\xi_{DL}^j$ of a PMA heterostructure also inversely correlates to $j_c$ in the spin-current-generating layer, i.e.,

$$j_c = (4e/\pi\hbar) \mu_0 M_s t_{FM} H_c / \xi_{DL}^j, \quad (6)$$

However, recent experiments [84] have shown that neither Equation (4) nor Equation (5) can provide a reliable prediction for the switching current and $\xi_{DL}^j$ and that there is no simple correlation between $\xi_{DL}^j$ and the critical switching current density of realistic perpendicularly magnetized spin-current generator/ferromagnet heterostructures and that SOT-driven switching of perpendicularly magnetized Hall bars in the micrometer or sub-micrometer scales cannot provide a quantitative estimation of $\xi_{DL}^j$ and the spin Hall ratio ($\theta_{SH}$), even though such switching experiments are of interest for advancing the understanding of magnetization reversal as well as domain wall depinning. The macrospin analysis simply does not apply to the switching dynamics of micrometer-scale samples so that the values of $\xi_{DL}^j$ determined using the switching current density and Equation (5) can produce overestimates by up to thousands of times (see $\xi_{DL,macro}^j$ and $\xi_{DL,macro}^j/\xi_{DL}^j$ in Table 2). [84]

A domain-wall depinning analysis based on Equation (6) can either under- or over-estimated $\xi_{DL}^j$ by up to tens of times (see $\xi_{DL,DW}^j$ and $\xi_{DL,DW}^j/\xi_{DL}^j$ in Table 2). [84] In addition, there is not any obvious connection between the discrepancy and the types, the coercivity, the anisotropy field, the thickness, and the magnetization of the materials. Caution is thus required to interpret data associated with perpendicular magnetization switching current density. For the same reason, so-called "switching efficiency", defined as the $H_c / j_{c0}$ ratio or the $H_k / j_{c0}$ ratio, can be highly misleading when used as a comparative guidance of $\xi_{DL}^j$ for different materials. This is particularly so when the switched FMs have significantly varying magnetic anisotropy, DMI, magnetic damping, layer thicknesses, thermal stability, pinning field, saturation magnetization, etc. Somewhat surprisingly, a perpendicular ferromagnetic $Co_{40}Fe_{40}B_{20}$ layer (thickness of 1.4 nm, coercivity of 15 Oe) has also been demonstrated to be switchable at a moderate charge current density of $4.2\times10^7$ A/cm² by a 4 nm Mo layer with a vanishingly small damping-like SOT efficiency of -0.003.[105] These observations consistently suggest that the switching current or "switching efficiency" of perpendicular heterostructures is not a quantitative indicator of $\xi_{DL}^j$.

**Table 2.** Comparison of spin-torque efficiencies determined from harmonic Hall voltage ($\xi_{DL}^j$) and magnetization switching ($\xi_{DL,DW}^j$, $\xi_{DL,macro}^j$) of perpendicularly magnetized bilayers. The values of $\xi_{DL,DW}^j$ are determined from a model of a current-induce effective field acting on domain walls [Equation (6)], and $\xi_{DL,macro}^j$ is determined within a macrospin model [Equation (5)]. The $\xi_{DL}^j$ results for the Pt 6 nm/Fe₃GeTe₂ 4 nm and Ta 5 nm/Tb₂₀Fe₆₄Co₁₆ 1.8 nm samples were reported in Refs. [137] and [128], while the corresponding values of $\xi_{DL,DW}^j$ and $\xi_{DL,macro}^j$ were calculated in Ref. [84] using the reported switching current density $j_c$ values and other sample parameters as reported in Refs. [137] and [128].

| Samples | $j_c$ ($10^7$ A/cm²) | $\xi_{DL}^j$ | $\xi_{DL,DW}^j$ | $\xi_{DL,macro}^j$ | $\xi_{DL,DW}^j/\xi_{DL}^j$ | $\xi_{DL,macro}^j/\xi_{DL}^j$ |
|---|---|---|---|---|---|---|
| Pt 2 nm/Co 1.4 nm (annealed) | 8.2 | 0.15 | 0.48 | 2.8 | 3.2 | 18.7 |
| Pt 4 nm /Co 0.75 nm | 3.2 | 0.21 | 0.23 | 6.0 | 1.1 | 28.6 |
| [Pt 0.6 nm/Hf 0.2 nm]₅/Pt 0.6 nm /Co 0.63 nm | 2.4 | 0.36 | 0.38 | 3.8 | 1.1 | 10.6 |
| Pt₀.₇₅Pd₀.₂₅ 4 nm /Co 0.64 nm | 2.6 | 0.26 | 0.36 | 4.8 | 1.4 | 18.5 |
| Au₀.₂₅Pt₀.₇₅ 4 nm /Co 0.64 nm | 1.7 | 0.30 | 1.05 | 5.2 | 3.5 | 17.3 |
| Pt₀.₇(MgO)₀.₃ 4 nm /Co 0.68 nm | 1.5 | 0.30 | 0.09 | 16.8 | 0.3 | 56 |
| Pd 4 nm /Co 0.64 nm | 3.75 | 0.07 | 0.09 | 3.0 | 1.3 | 42.9 |
| W 4 nm /Fe₀.₆Co₀.₂B₀.₂ 1.5 nm | 0.036 | 0.4 | 4.0 | 306 | 10 | 765 |
| Pt 6 nm /Fe₃GeTe₂ 4 nm | 1.2 | 0.12 | 0.06 | 2.0 | 0.5 | 16.7 |
| Ta 5 nm /Tb₂₀Fe₆₄Co₁₆ 1.8 nm | 0.04 | 0.12 | 2.1 | 234 | 18 | 1950 |

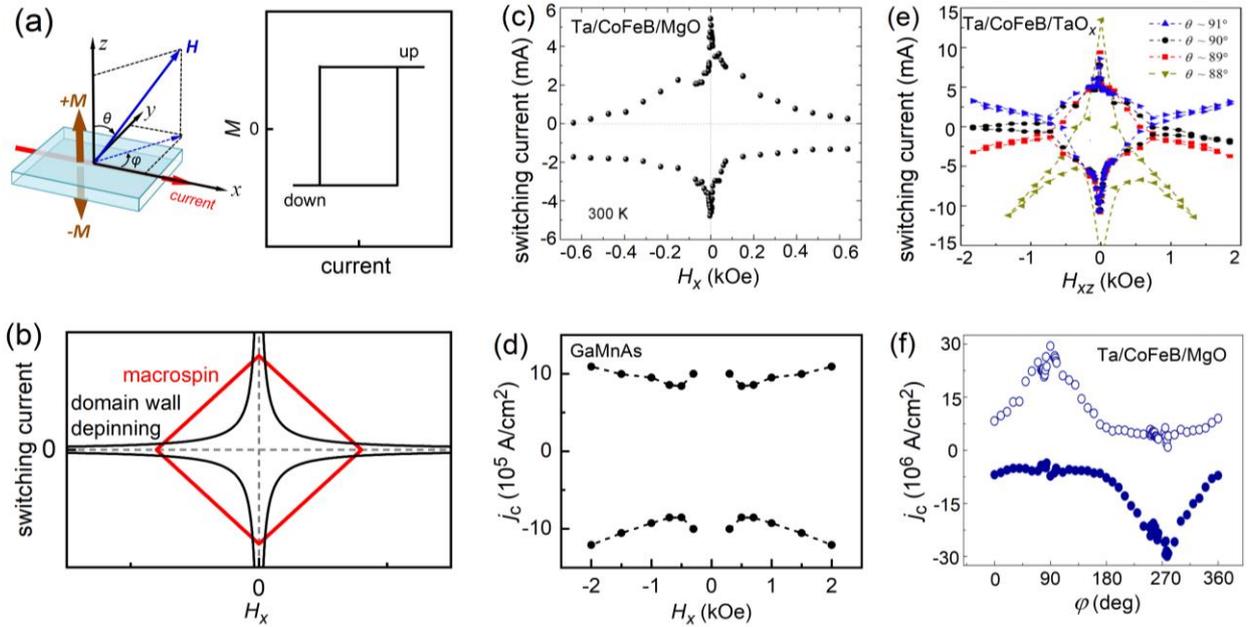

**Figure 10**. Qualitative deviation from the simplified macrospin and domain wall depinning models. (a) Schematic of configuration and coordinate of current-driven magnetization switching experiment. $\pm M$ represents the up and down states of the perpendicular magnetization, $\theta$ and $\varphi$ are the polar and in-plane angles of the applied magnetic field ($H$), and the charge current flows along the $x$ direction. Dependence on in-plane "assisting" magnetic field ($H_x$, collinear with the write current) of the switching current predicted (b) by the rigid macrospin model [Equation (5)] and by domain wall depinning model, suggesting that the switching currents for up-down and down-up magnetization reversal are of the same magnitude under the same $|H_x|$. (c) the switching current density ($j_c$) vs $H_x$ as measured for perpendicularly magnetized Ta (4 nm)/$Co_{40}Fe_{40}B_{20}$ (1 nm)/MgO (1.6 nm), indicating significant asymmetry in the magnitudes of $j_c$ for the same $|H_x|$. Reprinted with permission.[129] Copyright 2015, American Physical Society. d) $j_c$ vs $H_x$ as measured for perpendicularly magnetized $Ga_{0.94}Mn_{0.06}As$ (7 nm), suggesting an increase of the switching current density by a large $H_x$. Reprinted under a Creative Commons Attribution 4.0 International License.[46] Copyright 2019, Springer Nature. e) Switching current vs the applied magnetic field $H_{xz}$ (with the polar angle of 88°-91°) as measured for perpendicularly magnetized Ta (5nm)/$Co_{20}Fe_{60}B_{20}$ (1 nm)/$TaO_x$, suggesting that the switching current has a strong asymmetry, shift, increase, and strong tunability by a small perpendicular magnetic field. Reprinted with permission.[131] Copyright 2014, American Physical Society. f) $j_c$ vs $\varphi$ for Ta (4nm)/$Co_{40}Fe_{40}B_{20}$ (1 nm)/MgO (2 nm), indicating strong asymmetry in the switching density in presence of a transverse magnetic field $H_y$. Reprinted with permission.[132] Copyright 2019, American Physical Society.

Another puzzling observation is about the diversity of the effect on the switching current density of the in-plane magnetic fields (see the experimental configuration and coordinate in **Figure 10**a). For both the simplified macrospin [123,124] and domain wall models,[125-127] a transverse magnetic field ($H_y$) should play a minor role, while the longitudinal magnetic field ($H_x$, usually known as in-plane "assisting" field) should always aid the switching and reduce the switching current density at a same rate for positive and negative signs. In other words, the switching currents for up-down and down-up magnetization reversal are symmetric and have the same magnitude under the same $|H_x|$ (Figure 10b). This appears to be the case for many magnetic heterostructures (e.g., $Au_{1-x}Pt_x$/Co,[94] $Pd_{1-x}Pt_x$/Co,[81] W/$Fe_{60}Co_{20}B_{20}$,[84] Ta/TbFeCo,[128] $Pt_{0.7}(Si_3N_4)_{0.3}$/CoTb [95]), but not all. Instead, the switching current has been experimentally measured to be asymmetrically reduced by $H_x$ of opposite signs in some samples (e.g., Ta/$Co_{40}Fe_{40}B_{20}$/MgO [129] in Figure 10c or W/$Co_{40}Fe_{40}B_{20}$/MgO,[107,130] Hf/CoFeB,[107] $Ta_{48}N_{52}$/CoFeB/MgO.[107] In some other cases, $H_x$ can even hider current-driven switching of perpendicular magnetization and increase the switching current symmetrically or asymmetrically (e.g., the $Ga_{0.94}Mn_{0.06}As$ [46] and Ta/$Fe_{60}Co_{20}B_{20}$/$TaO_x$ samples [131] in Figure 10d,e). It has also been reported that a nonzero transverse field ($H_y$) can dramatically break the symmetry of the switching currents for up-down and down-up reversals (see Figure 10f and Ref. [132]), in contrast to the simplified macrospin and domain wall depinning models (Figure 7c). It is also puzzling that a small out-of-plane magnetic field component (or a tiny deviation of the polar angle from 90°) can dramatically alter the switching behavior of a perpendicular magnetization in some samples, e.g., the Ta/$Fe_{60}Co_{20}B_{20}$/$TaO_x$ samples [131,133] in Figure 10e.

The existence of these outstanding puzzles is consistent with the aforementioned inaccuracy of the existing macrospin and domain wall motion models in predicting the switching current density of PMA heterostructures. These studies together imply that important fundamental physics is still missing in the existing understanding of magnetization switching.

5.2 Influence of magnet dimension

Indeed, low switching current densities compared to the ones for state-of-art in-plane SOT-MRAM devices are possible for micrometer-sized devices, especially in the presence of the large long-duration current-induced thermal assist [e.g. $3.6 \times 10^5$ A/$cm^2$ for perpendicular W/$Fe_{60}Co_{20}B_{20}$ ($\xi_{DL}^j$ = 0.4, **Figure 11**a),[84] 4×10$^5$ A/cm$^2$ for perpendicular Ta/Tb$_{20}$Fe$_{64}$Co$_{16}$ ($\xi_{DL}^j$ = 0.12),[128] 2.8×10$^5$ A/cm$^2$ for perpendicular Bi$_2$Se$_3$/CoTb ($\xi_{DL}^j$ = 0.16),[134] and 6×10$^5$ A/cm$^2$ for in-plane Bi$_2$Se$_3$/Ni$_{81}$Fe$_{19}$ ($\xi_{DL}^j$ = 1.0),[135] 5.5×10$^6$ A/cm$^2$ for perpendicular FeTb single layer.[54] However, the write performance of devices in the regime of domain-wall-mediated switching is largely irrelevant to and cannot guide that of practical magnetic memory technologies. This is because the switching currents grow by orders of magnitude when the size of magnetic layers is reduced to the sub-30-nm scale where domain wall propagation is no longer possible (Figure 11b).[136] Consequently, it is not meaningful to simply compare the magnitudes of switching current density for devices of different magnetic dimensions or/and materials. Note that the structurally unstable nearly compensated ferrimagnets (e.g., CoTb) and very-low-anisotropy ferromagnets (e.g., Ni$_{81}$Fe$_{19}$) are also less likely to allow for fast reliable reading due to their inability to promote high-TMR MTJs.

5.3 Influence of Joule heating

During the process of spin-orbit-torque switching of perpendicular magnetization, Joule heating due to the large, quasi-dc switching current with a density of 10$^6$-10$^8$ A/cm$^2$ can be significant. When the driven perpendicular magnetization has very high Curie temperature and perpendicular magnetic anisotropy (such as some Pt/Co bilayers),[84] the Joule heating plays a relatively minor role in the switching process. However, for magnetic layers, e.g., ferrimagnetic alloys, sub-nanometer thick ferromagnets, van der Waals magnets, or diluted semiconductors, which have low Curie temperature, weak PMA, and/or low depinning field, the Joule heating can dramatically lower the switching barrier such that the apparent critical switching current density might be quite low. An interesting demonstration is that the temperature of a Pt 6 nm/Fe$_3$GeTe$_2$ 4 nm raises to be above the Curie temperature before the magnetization is switched by the current-induced spin-orbit torque [137] (Figure 11c). Another example is that the switching current of a strained Ga$_{0.94}$Mn$_{0.06}$As thin film [46] was reduced by a factor of 10 when the temperature is increased from 10 K to 70 K (Figure 11d).

From point of view of technology, the current-induced Joule heating does not assist the fast operation in the ns or sub-ns timescales. However, Joule heating can be significant in the case of a quasi-dc switching process. Care is needed in the characterizations of prototype devices of the perpendicular magnetization with low Curie temperature as well as significant Joule heating effect. For instance, there would be a significant alteration of the saturation magnetization and the magnetic anisotropy due to the Joule heating in the analysis of thermal factor and zero-temperature switching current density from ramp rate or current-pulse-width dependent measurements in the thermal-activated regime.

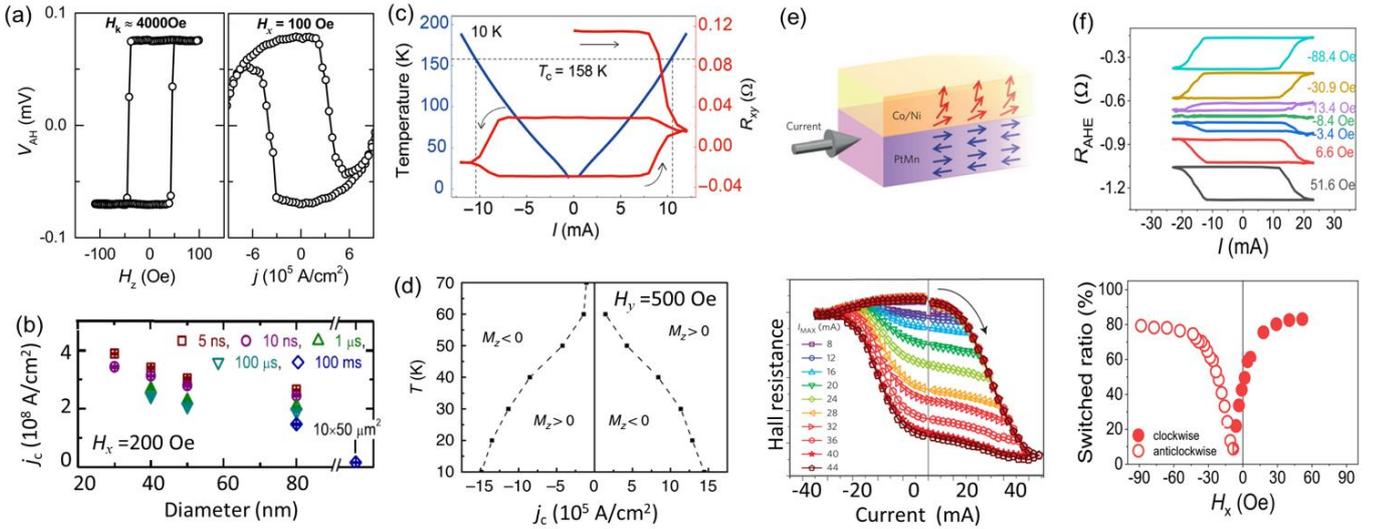

**Figure 11**. Variation of switching current density. a) Anomalous Hall voltage ($V_{AH}$) for perpendicularly magnetized W 4 nm/Fe$_{0.6}$Co$_{0.2}$B$_{0.2}$ 1.5 nm plotted as a function of out-of-plane magnetic field ($H_z$) and the density of charge current in the W layer ($j$), respectively. Reprinted with permission.[84] Copyright 2021, American Physical Society. b) Spin-orbit torque switching current density of magnetic tunnel junctions (MTJ) with different diameters (an in-plane longitudinal magnetic field of 200 Oe was applied, the different symbols represent different write current durations), indicating a dramatic increase with decreasing the MTJ diameter because of the gradual transition from multi-domain configuration to the macrospin configuration. Reprinted with permission.[136] Copyright 2015, AIP Publishing. c) Temperature and Hall resistance of a Pt 6nm/Fe$_3$GeTe$_2$ Hall bar vs the in-plane current, indicating significant Joule heating during the switching process. Reprinted under a Creative Commons Attribution NonCommercial License 4.0. [137] Copyright 2019, American Association for the Advancement of Science. d) Switching current density vs temperature ($T$) for a strained Ga$_{0.94}$Mn$_{0.06}$As with charge current flowing along the [-110] crystal axis. Reprinted under a Creative Commons Attribution 4.0 International License.[46] Copyright 2019, Springer Nature. e) Partial switching of perpendicularly magnetized PtMn (8 nm)/[Co (0.3 nm)/Ni (0.6 nm)]$_2$/Co (0.3 nm) with non-uniform exchange bias. The black arrow indicates the position from which the measurement starts. Reprinted with permission.[97] Copyright 2016, Springer Nature. f) Partial switching of Mn$_3$Sn (7 nm)/Cu (1 nm)/CoFeB (1.8 nm) with perpendicular magnetic anisotropy under different longitudinal fields ($H_x$). Reprinted under a Creative Commons Attribution 4.0 International License.[88] Copyright 2022, Springer Nature.

5.4 Influence of non-uniform switching

Current-induced partial switching of perpendicular magnetization is also commonly seen for the Hall bar devices in the literature. As shown in Figure 11e, the perpendicularly magnetized PtMn (8 nm)/[Co (0.3 nm)/Ni (0.6 nm)]$_2$/Co (0.3 nm) with interfacial exchange bias is not switched as an entirety, but the switched volume increases progressively as the switching current increases. [97] This suggests that these samples have a wide distribution of the energy barrier for switching by the spin-orbit torques. In this case, the switching current is also obviously a function of the ratio of the switched domains. Figure 11f shows a second example of the micrometer-sized Hall bar of Mn$_3$Sn (7 nm)/Cu (1 nm)/CoFeB (1.8 nm),[88] indicating partial switching with the switched ratio strongly tunable by the applied longitudinal magnetic field. These partial switching behaviors imply that some perpendicularly magnetized samples are not uniform, with some portions of the regions easier to switch by spin-orbit torque than others. Therefore, when quantitatively considering the switching current density, such partial switching characteristics should be taken into account. For example, the average values of coercivity, magnetization, and magnetic anisotropy field of the entire samples would be different from those of the switched portion of the sample.

5.5 Influence of current duration and switching probability

In addition to external magnetic fields, Joule heating, and the switching ratio, the critical switching current is also strongly dependent on the switching probability and the duration of the current pulse. **Figure 12**a,b shows the experimental data for an example sample of Pt (3 nm)/Co (0.6 nm)/AlO$_x$ with the perpendicularly magnetized Co layer patterned into a 95×95 nm$^2$ square dot.[57] Clearly, under a given in-plane assisting field $H_x$, the switching current density varies remarkably with the switching probability (Figure 12a). Meanwhile, for a given switching probability and an in-plane assisting field, the critical switching current is also dependent on the duration of the current.

The switching process is in the thermally activated regime when the current duration is much longer than the thermal attempt time (typically 1 ns). In this case, the critical current varies only weakly with the change in current duration. As shown in Figure 12b, this very weak dependence *happens* to reasonably fit the equation analytical developed for the thermally-activated macrospin model, [124] i.e.,

$j_c=(e\mu_0 M_s t_{FM} H_k/2\hbar\xi^j_{DL})\times[\pi-2H_x/H_k$
$-\sqrt{\frac{8}{\Delta}\ln\frac{-\tau_p}{\tau_0\ln(1-P)}-4(H_x/H_k)^2-4(\pi-4)H_x/H_k+\pi^2-8}]$  (7)

where $\tau_0$ is the thermal attempt time, $H_x$ the in-plane assisting magnetic field along the current direction, $H_k$ the perpendicular magnetic anisotropy field, $\Delta$ the energy barrier, and $P$ the switching probability (Figure 12a), respectively. the switching probability is dependent on the current duration, [124] i.e.,

$P = 1- \exp[-\tau_p/\tau_0 \exp(-\Delta/k_B T)]$,  (8)

where $k_B$ is the Boltzmann constant and $T$ the temperature. In qualitative agreement with Equation (7), the switching current density of a perpendicular magnetization is typically found to increase significantly at low write error rates. [57]

In the short-time regime, typically < 1 ns, where the current is too fast for the thermally-assisted switching, the switching current increases dramatically with the shortening duration of the current pulse. Specifically, $j_c$ scales inversely with $t_p$, [124] i.e.,

$j_c = j_\infty + q/\tau_p$  (9)

where $j_\infty$ is the intrinsic critical switching current and $q$ is a parameter related to the number of electrons that needs to be pumped into the system before a reversal occurs, describing the efficiency of angular momentum transfer from the current to the spin system. [57]

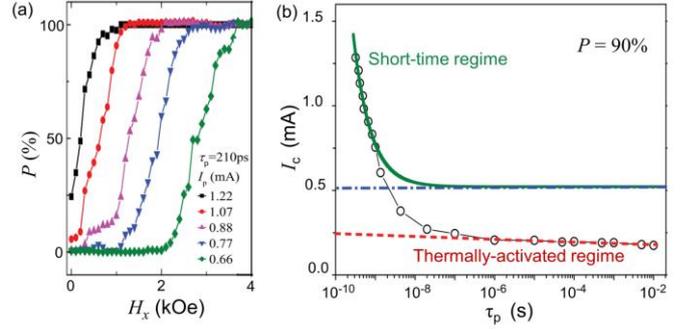

**Figure 12**. Fast switching of Pt (3 nm)/Co (0.6 nm)/AlO$_x$ sample with the Co layer patterned into a 95×95 nm$^2$ square dot. a) Switching probability ($P$) plotted as a function of in-plane assisting magnetic field ($H_x$) under different current amplitudes ($I_p$) but the same current duration ($\tau_p$) of 210 ps. b) Critical switching current at $P$ = 90% plotted as a function of $\tau_p$ measured with $H_x$ =0.91 kOe. Reprinted with permission.[57] Copyright 2014, AIP Publishing.

**6. Choice of perpendicular magnetic materials**

6.1. Ferromagnets vs ferrimagnet and antiferromagnets

In a switching process, both the spin-orbit torque and the switching barrier of the magnetization are critical. The switching barrier can be distinct for different types of materials, such as ferromagnets, ferrimagnets, and antiferromagnets (**Figure 13**a). As implied by Equation (4) and Equation (5), the switching barrier is in proportion to the saturation magnetization ($M_s$) in both macrospin and multi-domain regimes. The ferromagnetic metal CoFeB, which is favored for the high tunnel magnetoresistance of its MTJs, has relatively high $M_s$ of typically 700 - 1300 emu/cm$^3$, depending on the thickness and the growth protocol. Ferromagnetic metal Co, which is widely used partly for its good PMA in contact with the spin Hall metal Pt, also has a high $M_s$ of 1200-1400 emu/cm$^3$.

In contrast, ferrimagnetic metals (e.g., Fe$_x$Tb$_{1-x}$, [54] Co$_x$Tb$_{1-x}$, [134] (CoFe)$_x$Gd$_{1-x}$, and Mn$_x$Ga$_{1-x}$ [139,140]) can have very small or even negligible magnetization when the two antiferromagnetically aligned magnetic sub-lattices (Figure 13a) are tuned to be nearly compensated. This, at the first glance, seems to suggest an extremely low or even zero switching barrier for the nearly or fully compensated ferrimagnetic metals. However, the anisotropic field ($H_k$) and coercivity ($H_c$), which are relevant to the switching barrier of macrospin rotation ($\propto M_s H_k$) and domain wall depinning ($\propto M_s H_c$), respectively, also appear to increase rapidly (Figure 13b) or even diverge [141] at the magnetization compensation point of ferrimagnets. As a result, the vanishing saturation magnetization and diverging anisotropic field or coercivity near the compensation point make the exact barrier height of the ferrimagnetic metals and the prediction of the required switching current density open questions. Some experiments do show divergence of the required switching current at full compensation (see Figure 13e,f) for example on a Co/Ir/Co Hall bar which can be tuned to be fully compensated by the Ru thickness [142]). More efforts are required to clarify whether the compensated magnetic systems such as ferrimagnetic and antiferromagnetic metals had lower switching barriers for SOT applications.

Ideal antiferromagnets that have exactly zero magnetization would be a fully compensated ferrimagnet and thus are questioned as a low-barrier data bit material. Moreover, the fast, reliable electrical reading of antiferromagnets has remained a key challenge for practical applications. Note that while a nonzero anomalous Hall effect can be used for micrometer-sized or larger devices, the anomalous Hall detection is not scalable [143] and thus not a technologically useful readout for integrated circuits.

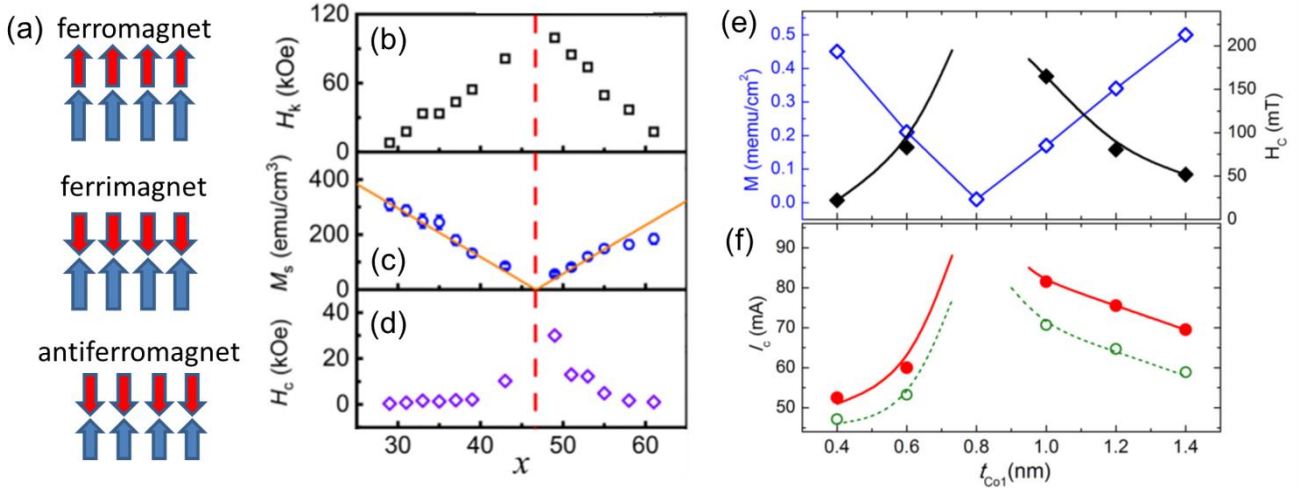

**Figure 13**. Switching energy barrier of ferrimagnets. a) Schematic of a ferromagnet, a ferrimagnet, and a collinear antiferromagnet, with the arrows indicating the alignment of the different magnetic sub-lattices. b) Perpendicular magnetic anisotropy field, c) Saturation magnetization, d) Coercivity of 20 nm $Fe_{100-x}Tb_x$ with different Tb concentration ($x$). b-d) Reprinted with permission.[54] Copyright 2022, AIP Publishing. SOT switching of Pt (5 nm)/Co1($t_{Co1}$)/Ir (0.55 nm)/Co2 (0.7 nm). e) Areal magnetic moment (blue), coercivity (black), and f) switching current under an in-plane longitudinal magnetic field of 1 kOe, as a function of Co1 thickness. The red solid circles and green open circles in f) indicate the current passing through the whole stack and the Pt layer, respectively. e,f) Reprinted with permission.[142] Copyright 2020, AIP Publishing.

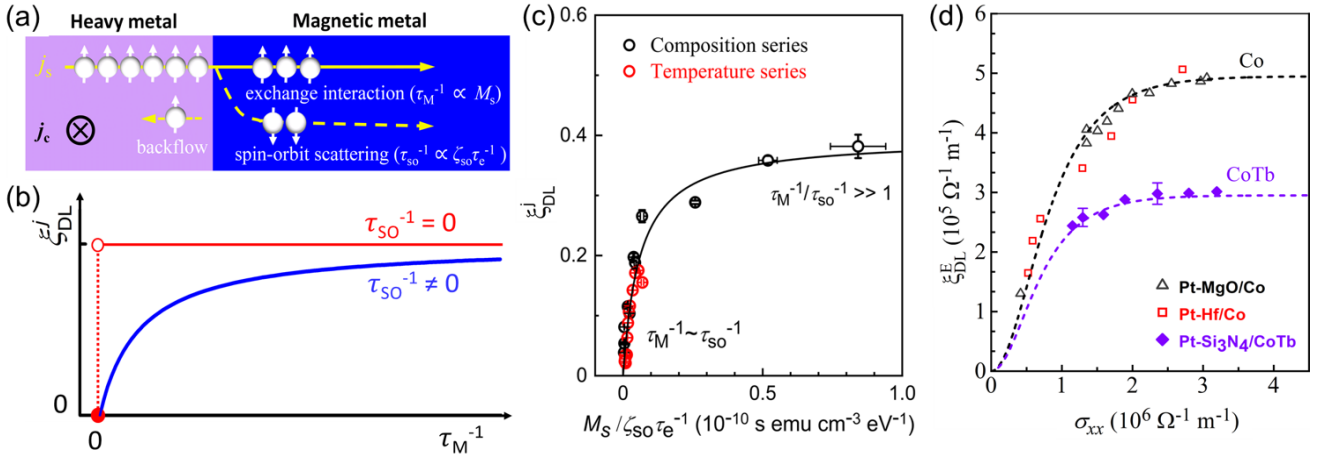

**Figure 14.** Variation of the spin-orbit torque with relative spin relaxation rates. a) Schematic of the spin relaxation processes that can influence the damping-like spin-orbit torque (SOT), highlighting the competition between exchange interaction (with relaxation rate $\tau_M^{-1} \propto M_s$) and spin-orbit scattering ($\tau_{so}^{-1} \propto \zeta_{so}\tau_e^{-1}$). Only the spin current relaxed by exchange interaction contributes to SOT. b) Dependence on $\tau_M^{-1}$ of the efficiency of the damping-like spin-orbit torque ($\xi_{DL}^j$) exerted by a given spin current on a magnetic layer with zero spin-orbit scattering ($\tau_{so}^{-1} =0$, red) and non-negligible spin-orbit scattering ($\tau_{so}^{-1} \neq 0$, blue), highlighting the critical role of spin-orbit scattering. c) $\xi_{DL}^j$ of $Pt_{0.75}Ti_{0.25}$ (6 nm)/$Fe_xTb_{1-x}$ (8 nm) vs $M_s/\zeta_{so}\tau_e^{-1}$ for the composition series ($x = 0.3-1$, $T = 300$ K, black circles) and for the temperature series ($x = 0.59$, $T = 25-300$ K, red circles). with under a Creative Commons Attribution 4.0 International License.[144] Copyright 2023, Springer Nature. d) Dependence on charge conductivity $\sigma_{xx}$ of the damping-like spin-orbit torque efficiency per electric field ($\xi_{DL}^E$) for Pt-MgO/Co, Pt-Hf/Co, and Pt-$Si_3N_4$/$Co_{0.65}Tb_{0.35}$. Despite the similar spin Hall ratio and interfacial spin transparency at each charge conductivity, the efficiency of the damping-like spin-orbit torque exerted on the $Co_{0.65}Tb_{0.35}$ is only 0.58 of that on Co. Reprinted with permission.[95] Copyright 2022, American Physical Society.

**Table 3**. Comparison of damping-like torque efficiency ($\xi_{DL}^j$) for FIMs and 3$d$ FMs in contact with spin current sources that have similar resistivities, thicknesses, and thus similar values of spin Hall ratio and interfacial spin transparency.

| spin current source | $\xi_{DL}^j$ | | ratio |
|---|---|---|---|
| | FIM | 3$d$ FM | |
| Ta | -0.03 (CoTb)[134] | -0.12 (FeCoB)[147] | 4 |
| W | -0.04 (CoTb)[148] | -0.44 (FeCoB)[7] | 11 |
| Pt | 0.017 (CoTb)[134] | 0.15 (Co,FeCoB)[147] | 8.8 |
| Pt/NiO | 0.09 (CoTb)[145] | 0.6 (FeCoB)[149] | 6.7 |
| Pt$_{0.75}$Ti$_{0.25}$ | 0.05 (Fe$_{0.5}$Tb$_{0.5}$)[144] | 0.38 (Fe)[144] | 7.6 |
| Bi$_2$Se$_3$ | 0.13 (GdFeCo)[146] | 3.5 (NiFe)[32] | 27 |

A more striking advance is that it has been very recently established that the SOT itself varies with the relative spin relaxation rates of the driven magnetic layer.[144] This is because, as shown in **Figure 14**a, spin currents relax within a magnetic layer not only via exchange interaction with the local magnetization (with relaxation rate $\tau_M^{-1} \propto M_s$) but also via spin-orbit scatting into the lattice (with relaxation rate $\tau_{so}^{-1} \propto \zeta_{so}\tau_e^{-1}$, $\zeta_{so}$ is the spin-orbit coupling strength, $\tau_e^{-1}$ is the electron scattering rate).[144] The dependence on the relative spin relaxation rates of damping-like SOT efficiency is given by [145]

$$\xi_{DL}^j \approx \xi_{DL,0}^j \, \tau_M^{-1}/(\tau_M^{-1}+ \tau_{so}^{-1}), \quad (10)$$

or

$$\xi_{DL}^j = T_{int}\theta_{SH} \, \tau_M^{-1}/(\tau_M^{-1}+ \tau_{so}^{-1}), \quad (11)$$

where $\xi_{DL,0}^j$ is the value of $\xi_{DL}^j$ in the limit $\tau_M^{-1}/\tau_{so}^{-1} \gg 1$, $\theta_{SH}$ is the spin Hall ratio of the spin current generator, and $T_{int}$ is the spin transparency of the interface between the spin current generator and the magnetic layer. The role of the relative spin relaxation rates was overlooked in the simple form of

$$\xi_{DL}^j = T_{int}\theta_{SH}. \quad (12)$$

Figure 14b schematically demonstrates the effect of spin-orbit scattering on SOTs suggested by Equation (10). Only in the case of a sufficiently thick magnetic layer with no spin-orbit scattering ($\tau_{so}^{-1} = 0$) and nonzero $M_s$, should the simple form of Equation (12) apply such that $\xi_{DL}^j$ is independent of $\tau_M^{-1}$ and $M_s$. This is a good approximation only for magnetic materials with $\tau_M^{-1}/\tau_{so}^{-1} \gg 1$, e.g., 3$d$ FMs that have high $M_s$, low resistivity, and weak SOC. However, in the presence of non-negligible spin-orbit scattering ($\tau_{so}^{-1} > 0$), $\xi_{DL}^j$ decreases more and more rapidly with reducing $\tau_M^{-1}/\tau_{so}^{-1}$ (with $\tau_M^{-1} \propto M_s$). This is generally the case of uncompensated "antiferromagnetic" domains ($M_s > 0$ emu/cm$^3$) and FIMs with strong SOC and large resistivities (e.g., Fe$_x$Tb$_{1-x}$ and Co$_x$Tb$_{1-x}$). However, $\xi_{DL}^j$ always diminishes at zero $\tau_M^{-1}$ ($M_s = 0$ emu/cm$^3$), e.g., in *perfectly* compensated ferrimagnets (FIMs) and antiferromagnets (AFs), and non-magnetic materials.

As shown in Figure 14c, the damping-like SOT efficiency of ferrimagnetic Pt$_{0.75}$Ti$_{0.25}$/Fe$_x$Tb$_{1-x}$ bilayers [144] is fit well by Equation (10) and varies monotonically with the relative spin relation rates in both composition- and temperature dependence studies. While spin-orbit scattering is weak in light 3$d$ ferromagnets (e.g., $\tau_M^{-1}/\tau_{so}^{-1} \gg 1$ for Fe, Co, Ni, CoFeB, etc.), it becomes an important or even the dominant mechanism of spin relaxation in ferrimagnets.[144] As compared in Table 3, for a given spin-current generating material, a partially compensated ferrimagnetic metal detects significantly weaker damping-like spin-orbit torque than its ferromagnetic counterpart. More examples can be found in Figure 14d. Despite the similar spin Hall ratio and interfacial spin transparency at each charge conductivity, the efficiency of the damping-like spin-orbit torque exerted on the Co$_{0.65}$Tb$_{0.35}$ is only 0.58 of that on Co. In fully compensated ferrimagnets or perfect collinear antiferromagnets, no effective damping-like spin-orbit torque or SOT switching can be expected.

6.2 Bulk SOT materials

Exerting a nonzero damping-like spin-orbit torque on a magnetic layer typically requires an external transverse spin current, while an in-plane charge current usually cannot generate any total spin-orbit torque or switching behaviors within centrosymmetric magnetic single layers (e.g., polycrystalline 3$d$ FeCoB [149,150] and Co [151]). Recently, experiments have established that a strong bulk-type net damping-like spin-orbit torque (**Figure 15**a) can be generated within single-crystalline GaMnAs with bulk inversion asymmetry [46] and some thick, strong spin-orbit coupling magnetic single layers, e.g., Co$_{1-x}$Pt$_x$, [52,114,115] Fe$_{1-x}$Pt$_x$, [53,120,121] Fe$_{1-x}$Tb$_x$, [54] Co$_{1-x}$Tb$_x$, [122,152] GdFeCo,[116,153] and $L1_0$-FeCrPt. [154] Switching of perpendicular magnetization by such bulk SOTs has been demonstrated. For example, Liu $et$ $al$. [54] reported bulk SOT switching of 20 nm Fe$_{0.42}$Tb$_{0.58}$ layer with a giant anisotropic field of 36.8 kOe, high coercivity of 1.72 kOe and moderate magnetization of 134 emu/cm$^3$ at a low current density of 5.5×10$^6$ A/cm$^2$ under an in-plane longitudinal magnetic field of ±3 kOe (Figure 15b). Such bulk SOT, particularly those in uniform magnetic layers, provides a new opportunity for the development of self-torqued, scalable memory and computing devices that are driven solely by the bulk SOT or by the combination of a bulk torque and an interfacial torque.

One of the remaining open questions is how to understand and optimize the spin current generation and inversion symmetry breaking associated with the bulk spin-orbit torque. First, recent experiments have unveiled that the bulk spin-orbit torque in the strong SOC centrosymmetric single layers arises from the interplay of a strong bulk spin Hall effect and an inversion symmetry breaking that alters the population balance of the spins of opposite polarizations. The bulk torque is found to be not subject to the presence of any composition gradient in some materials. [52,54,152] For instance, within the Fe$_{1-x}$Tb$_x$ and CoPt single layers, there was no observable vertical composition gradient within the resolution of scanning transmission electron microscopy and electron energy loss spectroscopy (Figure 15c,d). However, in some other materials, the presence of bulk SOT is accompanied by a composition gradient in others. [53,113-119,153] So far, it has remained unsettled as to the mechanism via which the composition non-uniformity affects the spin-orbit torque.

Furthermore, as a bulk effect, the bulk spin-orbit torque naturally has a strong dependence on the layer thickness of the magnetic layers. As shown in Figure 15e,f, it emerges and increases to very high efficiencies at large layer thicknesses (e.g., 300% for 66 nm $Fe_{0.57}Tb_{0.43}$ and 0.2 for 24 nm CoPt), but vanishes at small thicknesses. It would be interesting to advance the understanding of the mechanism that determines the increase rate of the bulk SOT with the layer thickness and the minimum thickness for a nonzero bulk SOT. This is because achieving very high SOT efficiency in magnetic films that have small thicknesses and high TMR is highly desirable for the development of high-performance SOT devices.

6.3 Spin-orbit torque superlattices

In conventional spin-current generator/magnet (SCG/M) bilayers (**Figure 16**a) the efficiency of the damping-like spin-orbit torque ($\xi_{DL}^j$) is usually low, particularly when $\theta_{SH}$ is small and when $T_{int}$ is far less than unity due to spin memory loss [153-161] and spin backflow [149,162-166] (e.g., $\xi_{DL}^j \approx$ 0.05-0.15 for Pt/3$d$ ferromagnet [16]). A more critical limitation of the bilayer scheme is that the efficiency of the damping-like SOT per magnetic layer thickness ($t$), $\xi_{DL}^j/t$, is inversely proportional to $t$. This degrades the effectiveness of the interfacial SOTs in spintronic applications where the M has to be thick to maintain a bulk [54,139,167] or shape [168,169] perpendicular magnetic anisotropy (PMA) and/or thermal stability when the lateral size scales down to tens of nanometers (e.g., magnetic tunnel junctions and racetrack nanowires).

Very recently, a new material scheme, spin-orbit superlattice, has been proposed and experimentally verified to remarkably boost SOT efficiency. [170] The main idea is to "accumulate" $\xi_{DL}^j$ by stacking SCG/M/oxide superlattice (**Figure 16**a). Here, the SCG generates a spin current that diffuses into the adjacent M, and the oxide breaks the inversion symmetry for a total torque to occur. In the SCG/M/oxide superlattice, the damping-like efficiency would be enhanced by a factor of the repeat number $n$, i.e., $\xi_{DL}^j = nT_{int}\theta_{SH}$, if all repeats are identical and the thickness of the M is greater than the spin dephasing length. When $n$ is infinite, the spin-orbit torque would, in principle, also be diverged. Any bulk SOT generated within the M layer is tentatively ignored here because bulk SOT typically vanishes at small thicknesses necessary for the superlattice scheme. Compared to the conventional SCG/M bilayer, the superlattice scheme [SCG/M/oxide]$_n$ can have greatly enhanced perpendicular magnetic anisotropy (Figure 16b), enhanced tunability, and $n^2$ times lower power consumption to generate a given spin torque strength than the corresponding SCG/M bilayer with the same total thicknesses for the SCG and the M.

The advantages of the superlattice [SCG/M/oxide]$_n$ have been verified using the prototype SCG Pt, the magnet Co, and the oxide MgO. [170] As shown in Figure 16b, the [Pt 2 nm/Co 0.8 nm/MgO 2 nm]$_n$ superlattices with large $n$ values exhibit very strong perpendicular magnetic anisotropy even when the corresponding single repeat (Pt 2 nm/Co 0.8 nm/MgO 2 nm) or the Pt/Co bilayer only have an in-plane magnetic anisotropy. Meanwhile, the value of $\xi_{DL}^j$ (Figure 16c) is also increasingly enhanced as the repeat number $n$ increases (note that $\xi_{DL}^j$ was not increased by a factor of $n$ in this specific sample due to degradation of structural quality in large-$n$ repeats, which can be avoided by improving the growth protocol). In Figure 16d, the spin-orbit torque of the superlattice [Pt 2 nm/Co 0.8 nm/MgO 2 nm]$_{10}$ switches the effectively 8 nm-thick perpendicular magnetic anisotropy Co ($H_k$=17.5 kOe, $H_c$= 0.6 kOe) at a current density of $\approx 2\times 10^7$ A/cm$^2$ within the Pt (the average current density of $\approx 1\times 10^7$ A/cm$^2$ for the entire superlattice). Note that switching of a 8 nm Co by interfacial spin-orbit torque of a bilayer scheme is unlikely to achieve.

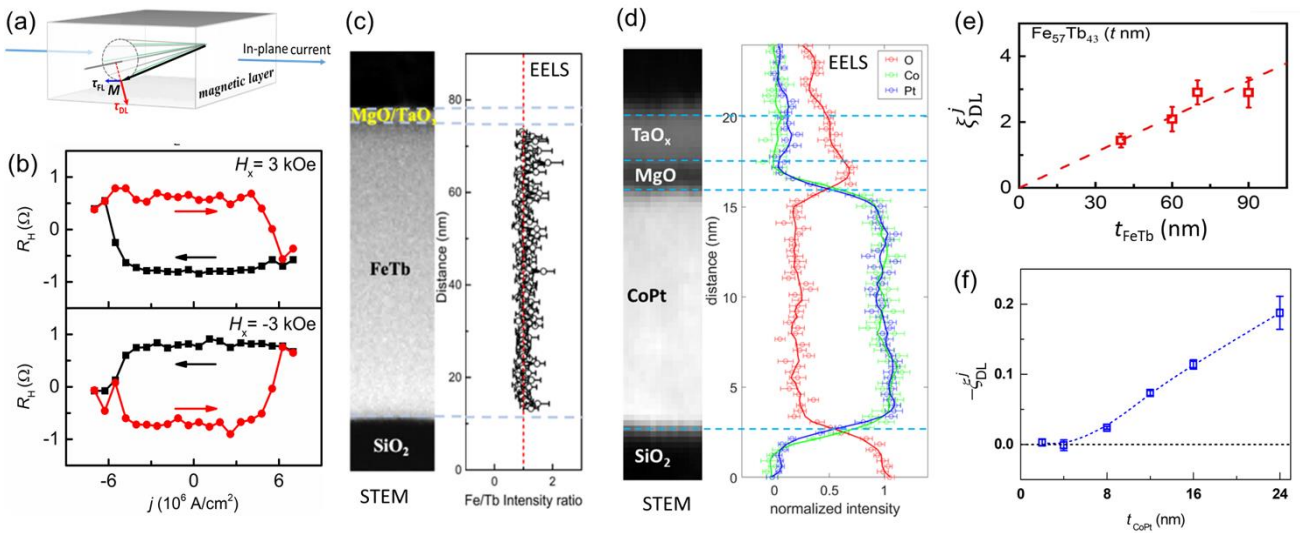

**Figure 15.** Bulk spin-orbit torque. a) Schematic of generation of bulk SOT by an in-plane current. b) Bulk SOT switching of 20 nm $Fe_{0.42}Tb_{0.58}$ layer ($M_s$ = 134 emu/cm$^3$, $H_k$ = 36.8 kOe, $H_c$ =1.72 kOe) under an in-plane longitudinal magnetic field of ±3 kOe. c) Scanning transmission electron microscopy (STEM) image and the electron energy loss spectroscopy (EELS) intensity ratio of Fe/Tb for a $Fe_{0.42}Tb_{0.58}$ sample. d) STEM image and the EELS intensity of Co, Pt, and O for a Si/SiO$_2$/CoPt (16 nm)/MgO (2 nm)/TaO$_x$. e) Layer thickness dependence of bulk damping-like torque efficiency for $Fe_{0.57}Tb_{0.43}$. f) Layer thickness dependence of bulk damping-like torque efficiency for CoPt. b,c,e) Reprinted with permission.[54] Copyright 2022, AIP Publishing. d,f) Reprinted with permission. [52] Copyright 2020, Wiley.

Such SCG/M/oxide superlattices provide new bricks for the development of energy-efficient, high-endurance, high-density SOT memory and computing technologies. The SCG/M/oxide material scheme should universally work for superlattices with the SCG being a spin Hall metal (not limited to Pt), an orbital Hall metal, a topological insulator, a complex oxide, and with the magnet being a ferromagnet, a ferrimagnet, or an antiferromagnet. This is in sharp contrast to the aforementioned bulk spin-orbit torque which is limited only to some specific materials. This novel [SCG/M/oxide]$_n$ material scheme might also be benefitted from using self-torqued magnetic materials as the M layers if the bulk SOT can be engineered to be significant at small thicknesses (still a challenge) and additive to the interfacial SOT. Finally, it is also important to note that no significant torque is expected or experimentally measured in the superlattice [SCG/M]$_n$/SCG with no symmetry-breaking oxide layer due to the cancellation of the spin currents from the two SCG layers sandwiching the M.

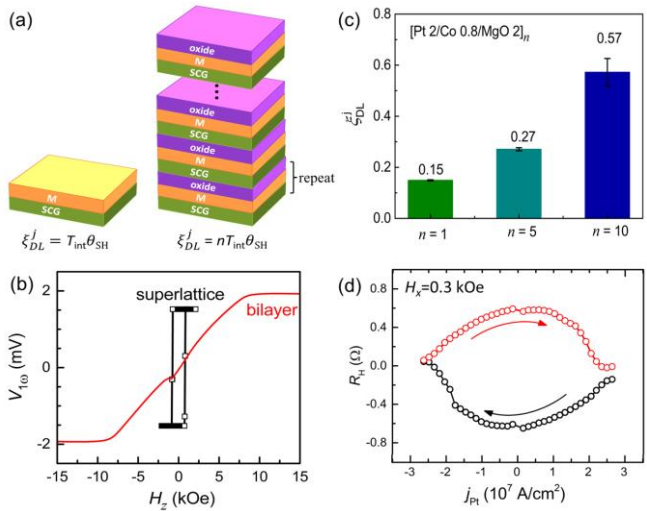

**Figure 16**. a) Damping-like spin-orbit torque efficiency expected for the spin current generator/magnet (SCG/M) bilayer and the [SCG/M/oxide]$_n$ superlattice. b) First harmonic Hall voltage vs the applied out-of-plane magnetic field for [Pt (2 nm)/Co (0.8 nm)/MgO (2 nm)]$_n$ superlattices with $n$ =1 and 5, indicating strong enhancement of perpendicular magnetic anisotropy in the superlattice scheme. c) Damping-like torque efficiency for [Pt (2 nm)/Co (0.8 nm)/MgO (2 nm)]$_n$ superlattices with $n$ =1, 5, and 10. d) Hall resistance of the superlattice [Pt (2 nm)/Co (0.8 nm)/MgO (2 nm)]$_{10}$ vs current density in the Pt, showing deterministic switching of perpendicular magnetic anisotropy Co (total Co thickness = 8 nm, perpendicular magnetic anisotropy field $H_k$=17.5 kOe, perpendicular coercivity $H_c$= 0.6 kOe, saturation magnetization $M_s$= 780 emu/cm$^3$) by the spin-orbit torque at a current density of ≈2×10$^7$ A/cm$^2$ within the Pt. An in-plane bias field of 0.3 kOe was applied along the current direction to overcome the DMI effect such that the spin torque can take effect. Reprinted with permission.[170] Copyright 2022, American Physical Society.

## 7. Prototype perpendicular SOT memory devices

The main motivation of research on switching of perpendicular magnetization by SOTs is to develop fast, dense, reliable low-power perpendicular SOT-MRAMs. The two major challenges for this goal are the requirements of a large in-plane assisting magnetic field and a large write current. Recently, there has been significant progress in the development of prototype perpendicular SOT-MRAM devices that can be switched without an external magnetic field.

K. Garello *et al*. [138] have demonstrated sub-nanosecond write of perpendicular SOT-MTJs in the absence of an external field by adding a Co hard magnet layer on the top of MTJs. The Co hard mask is used to provide the longitudinal assisting magnetic field, as patented first by Gaudin and collaborators in 2010. [171] **Figure 17**a shows a cross-sectional TEM image and switching current density of a 60 nm circular MTJ with a spin Hall channel of 3.5 nm W providing spin current and a 50 nm thick Co magnetic hard mask providing an in-plane field. While the switching of this prototype device is high (>100 MA/cm$^2$ for 1 ns duration of current pulses) and the Co hard magnetic mask is large in area (typically 170×390 nm$^2$) and require large gaps between each other, this design allows for optimization of the SOT-MTJ and the field component to improve the switching performance and scalability.

N. Sato *et al*. [112] and M. Wang *et al*. [111] have demonstrated field-free switching of perpendicular SOT-MRAMs via domain wall depinning by proper combination of SOT and STT. A schematic and pulse-width dependence of the switching current of such SOT+STT device are shown in Figure 17b. Grimaldi *et al*. [10] and Krizakova *et al*. [172-174] also demonstrated that the combination of SOT, STT, and the voltage control of magnetic anisotropy (VCMA) leads to reproducible sub-nanosecond switching with the speed of the cumulative switching time smaller than 0.2 ns. They also find from time-resolved measurements that SOT switching involves a stochastic two-step process that consists of domain nucleation and propagation, which have different genesis, timescales, and statistical distributions compared to STT switching.

It is important to emphasize that the perpendicular SOT-MTJ devices in the above demonstrations [10,111,112,138] were switched via domain nucleation and propagation. The switching current may increase further when these MTJs were scaled down for improved reliability and consistency (see Figure 11b). As discussed in Sect. 5, switching via domain nucleation and propagation has a much lower switching barrier than that via coherent rotation. So far, a demonstration of perpendicular SOT-MTJ devices that reverses as a macrospin is still missing. It is also worth mentioning that Kazemi *et al*. [175] proposed that a canted perpendicularly magnetized SOT-MRAM device, in which the long axis of an elliptic MTJ pillar is rotated by an angle with respect to the current direction, can be deterministically switched without the need for a magnetic field in both toggle and non-toggle modes using a damping-like SOT induced by an in-plane current pulse. This capability is realized by shaping the magnetic energy landscape as demonstrated using macrospin and micromagnetic simulations. Such a canted device is composed of layers with a uniform thickness and does not require the use of magnetic hard masks or any material other than those widely utilized in conventional spin-orbit devices. However, there has been no report on such perpendicular, canted SOT-MRAM devices although its in-plane counterpart has been demonstrated by S. Fukami *et al*. [176] and H. Honjo *et al*. [177]

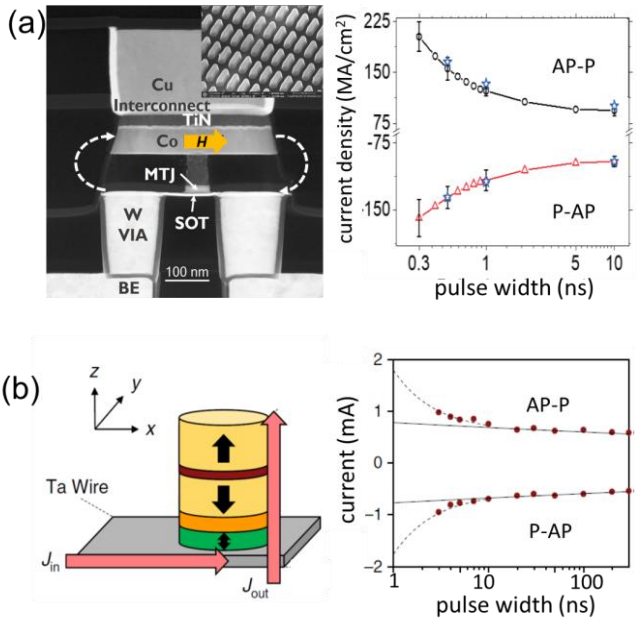

**Figure 17**. Protype perpendicular spin-orbit torque magnetic tunnel junctions (SOT-MTJs). a) Cross-sectional transmission electron microscopy image of a SOT-MTJ with 50 nm thick Co magnetic hard mask (inset is top scanning electron microscopy view) and the current density for 50% switching probability as a function of the pulse width. Reprinted with permission.[138] Copyright 2019, JSAP. b) Two-terminal SOT-MRAMs that utilize both SOT and spin-transfer torque (the spin Hall channel is made of Ta and 220 nm in width). Reprinted with permission.[112] Copyright 2018, Springer Nature.

## 8. Prototype perpendicular SOT logic devices

Spin-orbit torque switching of perpendicular magnetization can also be utilized to develop spintronic logic.[178-185] S. C. Baek et al.,[182] have proposed complementary Boolean logic operations based on spin-orbit torque and gate voltage-controlled magnetic anisotropy (VCMA, **Figure 18**a). When switched by an in-plane input clock current for a negative and positive gate voltage ($V_G$), Hall bars or MTJs based on Ta/CoFeB/MgO heterostructures can be defined as "p-type" and "n-type", a combination of which can function as the "XOR" and "AND" gates. There have also been reports on similar magnetic logic gates based on the perpendicular magnetization of ferrimagnets. Utilizing [Pt/Fe$_{1-x}$Tb$_x$/Si$_3$N$_4$]$_n$ multilayers, Dong et al.,[185] demonstrated reconfigurable multifunction in-memory logic features accompanying multistate memory states. Logic operatable without an external magnetic field has also been proposed.[183] Motion and switching of perpendicular magnetic anisotropy domains by spin-orbit torque and chiral coupling have also been demonstrated to enable "NOT", "NAND", and "NOR" gates (Figure 18b).[184] Based on SOT-driven magnetic domain wall motion in synthetic antiferromagnetic heterostructures, Wang et al.[186] have recently demonstrated spintronic leaky-integrate-fire spiking neurons with self-reset and winner-takes-all for neuromorphic computing. These advances in prototype devices shall stimulate future efforts on developing more energy-efficient perpendicular SOT logic devices.

## 9. Summary and Prospective

This review has discussed the advances and challenges in the electrical generation of spin current, the switching mechanisms and the switching strategies of perpendicular magnetization, the switching current density by spin-orbit torque of transverse spins, perpendicular magnetic materials, and memory and logic devices. This review has provided a systematic, in-depth understanding of physics, strategies, and devices of perpendicular magnetization switching by current-induced spin-orbit torques.

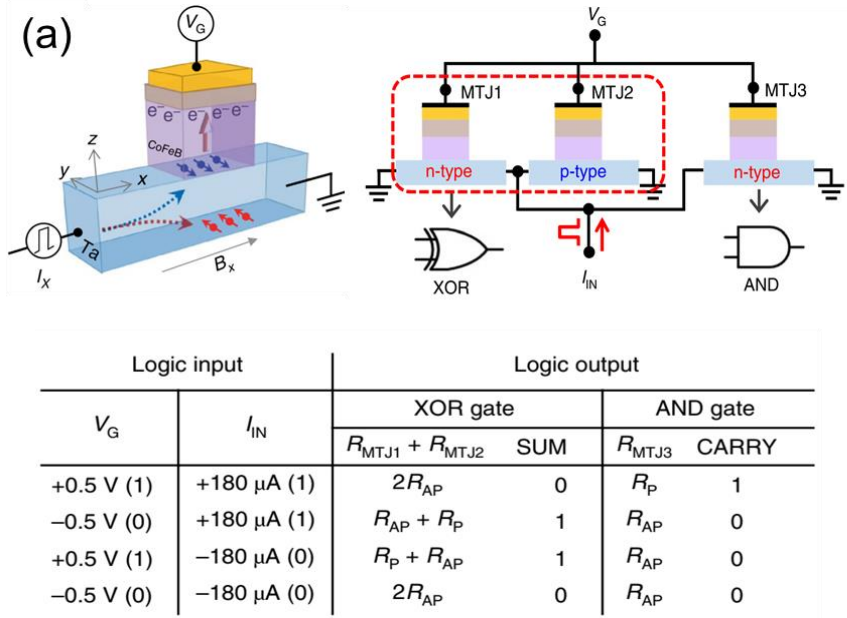

**Figure 18**. Protype perpendicular spin-orbit logic devices. a) Schematic and truth table of "XOR" and "AND" logic gates based on switching of perpendicular magnetic tunnel junctions (MTJs) by spin-orbit torque and voltage-controlled magnetic anisotropy. Reprinted with permission.[182] Copyright 2018, Springer Nature. b) Schematic of "NOT", "NAND", and "NOR" gates based on magnetic domain wall motion driven by spin-orbit torque and chiral coupling. Reprinted with permission.[184] Copyright 2020, Springer Nature.

Several critical questions remain open, which will have impacts on future development and material design.

(i) Despite the advances, field-free switching is only achieved for large SOT-MRAM devices that switch via domain nucleation and damping-like torque-driven domain wall depinning/propagation. Even so, the current density for full switching of the existing devices is still to be lowered. It is crucial to develop new strategies to achieve external-field-free switching of perpendicular magnetization in a low-current and scalable manner. It also remains interesting to explore new highly efficient source materials of spin currents of transverse, perpendicular, and/or longitudinal spins. Breakthroughs on this problem would readily benefit the development of compelling perpendicular SOT-MRAMs.

(ii) It is still a mystery as to what factors determine the critical switching current density of perpendicular magnetic materials. What is the exact quantitative correlation between the critical switching current density and the spin Hall ratio in heavy metal/ferromagnet bilayers? How are the DMI,[24,81,187,188] field-like torques, and magnetic damping involved in the nanosecond and sub-nanosecond SOT switching process of perpendicular magnetization?

(iii) Recent work has suggested switching of a large-area macrospin sample of Pt/Co/Cu/Ta by an in-plane charge current of 6 picoseconds in duration.[189] Can perpendicular SOT-MRAM devices also be switched so quickly? Is there a universal recipe to prepare micrometer-sized macrospin magnetic layers?

(iv) The switching current required for the reversal of perpendicular magnetization by SOTs is found to be asymmetric for positive and negative in-plane bias fields (Figure 10).[111,106,107,190] However, the underlying mechanism for this asymmetry remains a puzzle to clarify.

(v) While magnetically compensated materials are very interesting for their low magnetization and fast dynamics,[191-193] it remains unknown as to whether SOT-MRAM technology can benefit from compensated magnetic free layers, such as ferrimagnets and antiferromagnets. Is it possible to develop thermally stable, high-tunnel-magnetoresistance MTJs utilizing ferrimagnets or synthetic ferrimagnets (e.g., CoFeB/Gd/CoFeB [194])? Can the fast dynamics of the antiferromagnetic coupling systems provide a new opportunity for SOT-MRAM devices that have dimensions of a few tens of nanometers? What is the optimal trade-off between different critical parameters?

(vi) Bulk-type spin-orbit torque in uniform magnetic layers [52,54] in a few to tens nanometer thickness range provides a new driving force for magnetization switching in good scalability. The remaining challenge is to understand and optimize the associated symmetry breaking to generate very strong SOTs in magnetic thin films that have small thicknesses and provide high TMR.

(vii) Field-free magnetization switching by the in-plane current was attributed to the presence of an out-of-plane composition gradient in some literature.[114-116] However, some other samples with significant composition gradients [120-122] cannot be switched without an external field. If the composition was indeed the cause of field-free switching, what is the explanation for this difference? If not, what is the true underlying mechanism?

(viii) Contrary to the expectation from the widely assumed macrospin or domain wall motion models, some works have indicated that an in-plane field collinear with current may hider the switching process of some materials (e.g., GaMnAs,[46] etc.), increasing the switching current density with increasing in-plane bias field. How does the spin-orbit torque interplay exactly with the DMI? Any correlation to the partial switching? Unveiling the in-depth mechanism of this unexpected hider effect of the so-called assisting field may significantly advance the understanding of switching phenomena of perpendicular magnetization by spin-orbit torque.

(ix) Perpendicularly magnetized synthetic antiferromagnets (SAFs) (e.g., ferromagnet/Ru or Ir/ferromagnet or ferromagnet/Ru or Ir/ferrimagnet) can integrate the merits of ferromagnets and antiferromagnets. Specifically, SAFs can have enhanced thermal stability, low stray field, compensated magnetization, tunable interlayer exchange coupling via the Ruderman-Kittel-Kasuya-Yosida interaction and DMI,[195,196] external-field-free magnetization switching by SOTs,[197,198] etc. SAFs can thus be potentially compelling information storage layers of magnetic memories, logic gates, or neurons.[186] On the other hand, SAFs should also have an increased switching barrier (see Figure 13e for coercivity of the Co/Ir/Co) and reduced SOT efficiency (Sec. 6.1 and Figure 14) upon approaching full magnetization compensation. It is crucial to determine how to best benefit magnetic memory and computing technologies utilizing SAFs and SOTs.

(x) The current study of van der Waals magnets is mainly focused on the synthesis and basic magnetism and electrical transport characterizations. With future breakthroughs in developing air-stable, high-Curie-temperature, strong-magnetic-anisotropy, large-area, thin-film van der Waals magnets, integration of van der Waals magnets into spintronics is becoming a new blooming research frontier. From point of view of spintronics, there are many interesting open questions regarding van der Waals magnets. How is the spin transport different at the interface or in the bulk of van der Waals magnets compared to the well-studied metallic and oxide thin-film magnets? Can the strong tunability and the van der Waals interface significantly benefit the practical spintronic devices (magnetic memory, logic, and sensors)?

The progress we have described to understand and improve the switching performance of perpendicular magnetization by the SOTs suggests a promising future. Fully electrical switching of perpendicular magnetization with good stability and scalability by the SOTs of various spin polarizations can be a very compelling basis for enabling energy-efficient, fast, reliable, dense magnetic memory and computing technologies in the future. Considerable experimental and theoretical efforts are expected to be stimulated by the crucial open questions and the key challenges (e.g., the high switching current density, requirement of an effective in-plane field for switching by transverse spins, lack of complete switching of perpendicular magnetization, and low efficiency of electrical generation of perpendicular spins and longitudinal spins). Breakthroughs in these topics might trigger mass production of the non-volatile perpendicular SOT-MRAMs (1-2 transistors) as non-volatile SRAM replacement (6 transistors per bit) in high-performance computing systems. Note that sub-nanosecond write of in-plane SOT-MRAMs by CMOS transistors [177] and superconducting transistors [199,200] have been demonstrated. In addition, SOT-driven gradual magnetization switching and magnetic domain wall motion [186] may enable the development of novel beyond-CMOS functionalities, e.g., artificial synapses and neurons, [201] for neuromorphic computing in the field of artificial intelligence (AI) and internet of things (IoT).

**Acknowledgments**
This work is supported partly by the National Key Research and Development Program of China (Grant No. 2022YFA1200094), partly by the National Natural Science Foundation of China (Grant No. 12274405), and partly by the Strategic Priority Research Program of the Chinese Academy of Sciences (XDB44000000).

**Conflict of Interest**
The author declares no conflict of interest

**Keywords:** Spin-orbit torque, magnetic anisotropy, spin polarization, spin current, magnetization switching